\documentclass[aps,twocolumn,a4,showkeys,10pt,longbibliography,superscriptaddress]{revtex4-1}

\usepackage{graphicx}
\usepackage[english]{babel}
\usepackage{amsmath}
\usepackage{float}
\usepackage{hyperref}
\usepackage{subfigure}
\usepackage{tikz}
\usepackage{physics}
\usepackage{breqn}
\usepackage{mathtools}
\usepackage{enumerate}
\usepackage{amssymb}
\usepackage{bbold}

\begin{document}

\title{Quantum Fokker-Planck Master Equation for Continuous Feedback Control}
\date{\today}

\author{Bj\"{o}rn Annby-Andersson}
\email{bjorn.annby-andersson@teorfys.lu.se}
\affiliation{Physics Department and NanoLund$,$ Lund University$,$ Box 118$,$ 22100 Lund$,$ Sweden.}

\author{Faraj Bakhshinezhad}
\affiliation{Physics Department and NanoLund$,$ Lund University$,$ Box 118$,$ 22100 Lund$,$ Sweden.}

\author{Debankur Bhattacharyya}
\affiliation{Institute for Physical Science and Technology$,$ University of Maryland$,$ College
Park$,$ MD 20742$,$ USA.}

\author{Guilherme De Sousa}
\affiliation{Department of Physics$,$ University of Maryland$,$ College Park$,$ Maryland 20742$,$ USA.}

\author{Christopher Jarzynski}
\affiliation{Institute for Physical Science and Technology$,$ University of Maryland$,$ College
Park$,$ MD 20742$,$ USA.}

\author{Peter Samuelsson}
\affiliation{Physics Department and NanoLund$,$ Lund University$,$ Box 118$,$ 22100 Lund$,$ Sweden.}

\author{Patrick P. Potts}
\affiliation{Physics Department and NanoLund$,$ Lund University$,$ Box 118$,$ 22100 Lund$,$ Sweden.}
\affiliation{Department of Physics$,$ University of Basel$,$ Klingelbergstrasse 82$,$ 4056 Basel$,$ Switzerland.}

\begin{abstract}
Measurement and feedback control are essential features of quantum science, with applications ranging from quantum technology protocols to information-to-work conversion in quantum thermodynamics. Theoretical descriptions of feedback control are typically given in terms of stochastic equations requiring numerical solutions, or are limited to linear feedback protocols. Here we present a formalism for continuous quantum measurement and feedback, both linear and nonlinear. Our main result is a quantum Fokker-Planck master equation describing the joint dynamics of a quantum system and a detector with finite bandwidth. For fast measurements, we derive a Markovian master equation for the system alone, amenable to analytical treatment. We illustrate our formalism by investigating two basic information engines, one quantum and one classical. 
%We derive a general formalism for continuous feedback control in quantum systems, including the effect of a detector with finite bandwidth. Our main result is an equation describing the joint system-detector dynamics. For fast measurements, we derive a Markovian master equation for the system alone which can describe nonlinear feedback protocols. We illustrate our results with two simple examples, one quantum and one classical. Our formalism is promising for describing Maxwell demon type scenarios and illuminates the connection between quantum thermodynamics and information theory.
\end{abstract}

\maketitle

\textit{Introduction}. Quantum measurement and feedback control are key elements for emerging quantum technologies, enabling a wide range of applications, including quantum error correction \cite{Sarovar-PRA-2004}, deterministic entanglement generation \cite{Riste-Nature-2013}, atomic clocks \cite{Ludlow-RevModPhys-2015}, and quantum state stabilization \cite{Smith-PRL-2002,Sayrin-Nature-2011,Vijay-Nature-2012}. The last two decades have also witnessed a large number of fundamental experiments on feedback control of quantum systems \cite{Armen-PRL-2002, D'Urso-PRL-2003, Bushev-PRL-2006, Higgins-Nature-2007, Gillett-PRL-2010, Wheatley-PRL-2010, Xiang-NatPhot-2011, Zhou-PRL-2012, Okamoto-PRL-2012, Riste-PRL-2012, Yonezawa-Science-2012,Minev-Nature-2019}. Of special interest are experiments in quantum thermodynamics \cite{Vinjanampathy-Contem.Phys-2016} -- by using measurement and feedback, processes that are otherwise forbidden by the second law of thermodynamics may be realized, compellingly illustrated by Maxwell's demon \cite{MaxwellTheory_of_heat-book, Maxwell-demon-2-book, Maruyama-RMP-2009}. Over the last ten years, the demon has been realized in a wide range of experimental settings, both in classical \cite{Serreli-Nature-2007, Toyabe-Nat.Phys-2010, Koski-PNAS-2014, Koski-PRL-2014, Chida-Nat.Com.-2017, Kumar-Nature-2018, Barker-FDR-2021} and, recently, quantum systems \cite{Vidrighin-PRL-2016, Cottet-PNAS-2017, Masuyama-Nat.Com.-2018, Naghiloo-PRL-2018, Ribezzi-Nat.Phys.-2019}. This activity has inspired further work investigating the connection between thermodynamics and information theory \cite{Sagawa-Prog.Theor.Phys-2012, Parrondo-Nat.Phys-2015, Goold-J.Phys.A-2016}, and has resulted in generalizations of the second law for feedback controlled systems \cite{Sagawa-PRL-2008, Sagawa-PRL-2010, Ponmurugan-PRE-2010, Horowitz-PRE-2010, Morikuni-J.Stat.Phys-2011, Sagawa-PRL-2012, Sagawa-PRE-2012, Abreu-PRL-2012, Funo-PRE-2013, Wachtler-New.J.Phys-2016, Potts-PRL-2018}. A promising platform for exploring feedback control within quantum thermodynamics is solid state electronic systems \cite{Pekola-Nat.Phys-2015}, ranging from semiconductor quantum dots \cite{vanderWiel-RevModPhys-2002} to superconducting qubits \cite{Kjeargaard-Annu.Rev.Condens.Matter.Phys-2020}. Key features in these systems are large and fast tunability of system properties \cite{Fasth-NanoLetters-2005,GoranJohansson-book,Barker-APL-2019} and time resolved measurements \cite{Kung-PRX-2012,Hofmann-PSSb-2017}. Moreover, both discrete \cite{Riste-PRL-2012-2,Campagne-Ibarcq-PRX-2013,Barker-FDR-2021} and continuous \cite{Vijay-Nature-2012,Chida-Nat.Com.-2017} feedback protocols have been demonstrated experimentally.

The theoretical description of feedback control in quantum systems is typically based on stochastic differential equations \cite{Belavkin-ARC-1983, Belavkin-book-1987, Belavkin-JMA-1992, Belavkin-Com.Math.Phys-1992, Wiseman-PRL-1993, Wiseman-PRA-1994, Yanagisawa-book-1999, Doherety-PRA-1999, Korotkov-PRB-2001, Wiseman/Milburn-book-2010,Jacobs-book-2014,Zhang-Phys.Rep-2017} -- powerful tools that can describe discrete as well as continuous feedback protocols. In general, these equations must be solved numerically, providing limited qualitative insight. An important exception, amenable to analytical treatment, is the Wiseman-Milburn equation \cite{Wiseman-PRL-1993}, a Markovian master equation for continuous feedback protocols that depend linearly on the measured signal. However, optimal control often requires nonlinear protocols, for instance bang-bang control \cite{Kirk-Optimal-control-book, Cavina-PRA-2018} which has promising thermodynamic applications in solid state architectures \cite{Schaller-PRB-2011, Averin-PRB-2011, Chida-Nat.Com.-2017, Annby-PRB-2020}. For such continuous, nonlinear feedback protocols, no master equation description exists, emphasizing a need for further analytical tools. We stress that the word "nonlinear" here refers to the protocol's dependence on the measured signal, not to the system's dynamics.

\begin{figure}[H]
\centering
\includegraphics[scale=0.8]{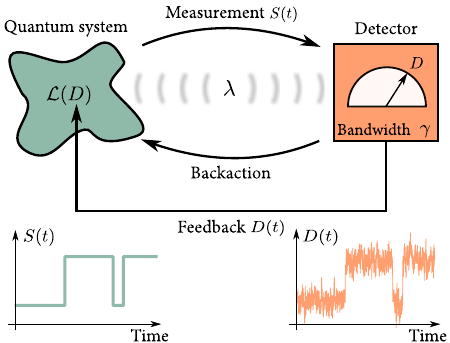}
\caption{Illustration of a generic measurement and feedback setup, consisting of an open quantum system and a detector with finite bandwidth $\gamma$. The detector continuously measures an arbitrary system observable. The measurement strength $\lambda$ determines measurement backaction. Continuous feedback is applied using the measurement outcome $D$ to control the Liouville superoperator $\mathcal{L}(D)$ of the system. The time traces visualize trajectories for the system state $S(t)$ and the measurement record $D(t)$.}
\label{fig:general-setting}
\end{figure}

In this letter, we satisfy this need by developing a general framework for continuous measurement and feedback control in quantum systems, able to provide analytical insight into nonlinear feedback protocols. Our main result, Eq.~(\ref{eq:MAIN-RESULT}) below, is a quantum Fokker-Planck master equation describing the joint dynamics of a quantum system and a detector with finite bandwidth (see Fig.~\ref{fig:general-setting}). This

\noindent equation is applicable to any quantum or classical system undergoing continuous feedback control. For fast measurements, Eq.~(\ref{eq:MAIN-RESULT}) reduces to a Markovian master equation for the system alone, generalizing the Wiseman-Milburn equation to nonlinear feedback protocols. The broad scope of Eq.~(\ref{eq:MAIN-RESULT}) suggests that our results will impact a wide variety of topics where nonlinear, continuous feedback control can be applied, such as quantum error correction \cite{Sarovar-PRA-2004}, entanglement generation \cite{Riste-Nature-2013}, quantum state stabilization \cite{Vijay-Nature-2012}, Maxwell's demon \cite{Averin-PRB-2011,Annby-PRB-2020} and machine learning \cite{Porotti-arXiv-2021}.

To illustrate our formalism, we investigate two toy models, a classical and a quantum two-level system, operated via nonlinear feedback protocols. For the classical model, we also derive a fluctuation theorem, highlighting the role of continuous measurement and feedback in information thermodynamics.

%This broad scope suggests that our results will have an impact on a wide variety of topics where nonlinear, continuous feedback control is important, such as \cite{Porotti-arXiv-2021}. (error correction, machine learning, entanglement generation, state stabilization)

\textit{Fokker-Planck master equation.} A general setup for continuous measurement and feedback is depicted in Fig.~\ref{fig:general-setting}. We consider an open quantum system whose dynamics, in the absence of measurement and feedback, are described by a Liouville superoperator $\mathcal{L}$. A detector continuously measures a system observable $\hat{A}$. The measurement strength $\lambda$ determines the magnitude of the measurement backaction, the limit $\lambda \rightarrow 0$ ($\lambda \rightarrow \infty$) corresponds to a weak, non-intrusive (strong, projective) measurement preserving (destroying) the quantum coherence of the system. Weak measurements thus reduce backaction, but increase measurement uncertainty. To provide a realistic detector description, we consider a finite bandwidth $\gamma$, acting as a low-pass frequency filter, eliminating high frequency measurement noise at the cost of introducing a time delay scaling as $1/\gamma$. Feedback control is incorporated by continuously feeding back the measurement outcome $D$ into the system, controlling the system Liouville superoperator via $\mathcal{L}(D)$.

Our main result is the following deterministic Fokker-Planck master equation (derivation outlined below), 
\begin{equation}
\begin{aligned}
\partial_t \hat{\rho}_t(D) &= \mathcal{L}(D) \hat{\rho}_t(D) + \lambda \mathcal{D}[\hat{A}] \hat{\rho}_t(D) \\
&- \gamma \partial_D \mathcal{A}(D) \hat{\rho}_t(D) + \frac{\gamma^2}{8 \lambda} \partial_D^2 \hat{\rho}_t(D),
\end{aligned}
\label{eq:MAIN-RESULT}
\end{equation}
\noindent describing the joint system-detector dynamics under continuous measurement and feedback control. The density operator $\hat{\rho}_t(D)$ represents the joint state of system and detector, where $\hat{\rho}_t \equiv \int dD \hat{\rho}_t(D)$ is the system state for an unknown measurement outcome $D$, and $P_t(D) \equiv \trace\{\hat{\rho}_t(D)\}$ defines the probability distribution of the measurement outcome $D$. Note that $\int dD P_t(D) = 1$ and $\trace\{\hat{\rho}_t\}=1$, see Supplemental Material (SM) below. The first term on the RHS of Eq.~(\ref{eq:MAIN-RESULT}) describes the feedback-controlled evolution of the system. This term allows for feedback protocols that are nonlinear in $D$. The second term, where $\mathcal{D}[\hat{A}]\hat{\rho} = \hat{A}\hat{\rho}\hat{A}-\frac{1}{2}\{\hat{A}^2,\hat{\rho}\}$ (note $\hat{A}^\dagger=\hat{A}$) describes how the system is dephased in the eigenbasis of $\hat{A}$ at a rate proportional to $\lambda$ due to measurement backaction. The last two terms constitute a Fokker-Planck equation describing the detector time evolution. These terms define an Ornstein-Uhlenbeck process \cite{Gardiner-book-stochastic-methods} with a system dependent superoperator drift coefficient $\mathcal{A}(D)\hat{\rho} \equiv \frac{1}{2}\{ \hat{A} - D,\hat{\rho} \}$ and diffusion constant $\gamma/8\lambda$. This describes a noisy relaxation of the measurement outcome towards a value determined by the system state. The derivation of Eq.~(\ref{eq:MAIN-RESULT}) is rather involved, see details in SM. The main text instead aims to highlight its implications and applications. However, we sketch the derivation at the end of the letter.

Equation (\ref{eq:MAIN-RESULT}) is, like most formalisms for continuous measurement and feedback, typically restricted to numerical solutions. However, when there exists a wide separation between the system and detector timescales, Eq.~(\ref{eq:MAIN-RESULT}) simplifies to a Markovian master equation for the system state $\hat{\rho}_t$, allowing for analytical treatment. The detector timescale $1/\gamma$ appears in the last two terms in Eq.~(\ref{eq:MAIN-RESULT}), and the system timescale $1/\Gamma$ is determined by $\mathcal{L}(D)+\lambda\mathcal{D}[\hat{A}]$. The role of $\lambda$, the measurement strength, is subtle, see below. When $\gamma \gg \Gamma$, $\hat{\rho}_t$ evolves, to first order in $1/\gamma$, according to
%
%To this end, we assume that the detector timescale $1/\gamma$ is much smaller than the dominating system timescale $1/\Gamma$, i.e., the time scale characterizing the first two terms on the right hand side of Eq.~(\ref{eq:MAIN-RESULT}), such that $\gamma \gg \Gamma$.  Under this condition, the system state evolves, to first order in $1/\gamma$, according to the Markovian master equation
%
\begin{equation}
\partial_t \hat{\rho}_t = \left[ \mathcal{L}_0 + \lambda \mathcal{D}[\hat{A}] + \gamma^{-1} \mathcal{L}_{\text{corr}}\right] \hat{\rho}_t,
\label{eq:Markovian-ME}
\end{equation}
\noindent with zeroth order Liouville superoperator $\mathcal{L}_0$ and first order correction $\mathcal{L}_{\text{corr}}$. $\mathcal{L}_0$ is obtained by approximating the system-detector density operator as $\hat{\rho}_t(D) = \left[ \sum_{aa'} \pi_{aa'}(D) \mathcal{V}_{aa'} \right] \hat{\rho}_t$, with
\begin{equation}
\pi_{aa'}(D) = \sqrt{4\lambda/\pi \gamma} e^{-(4\lambda/\gamma)[D - (\xi_a+\xi_{a'})/2]^2 },
\label{eq:stationary-OU-distr}
\end{equation}
and superoperators $\mathcal{V}_{aa'}\hat{\rho}\equiv \mel{a}{\hat
\rho}{a'}\dyad{a}{a'}$, where we used the eigenvalues and eigenvectors of the measured operator $\hat{A}=\sum_a \xi_a \dyad{a}$.  In this approximation, the detector is always in a system dependent stationary distribution $\pi_{aa'}(D)$. This is justified for $\gamma\gg\Gamma$, where changes of the system occur with a rate much smaller than the inverse detector relaxation time. Inserting this approximation in Eq.~(\ref{eq:MAIN-RESULT}) results in $\mathcal{L}_{0} =  \int dD \mathcal{L}(D) \left[ \sum_{aa'} \pi_{aa'}(D) \mathcal{V}_{aa'} \right]$, describing the system dynamics for a detector with zero delay time. The first order correction $\gamma^{-1}\mathcal{L}_{\rm corr}$ accounts for the lag of the detector due to its finite response time $\gamma^{-1}$. As usual in linear response theory, this correction can be written in terms of time-integrated correlation functions -- see SM. Note that $\lambda$ plays a special role in the separation of timescales since it appears both in the first and second line of Eq.~(\ref{eq:MAIN-RESULT}). In general, Eq.~(\ref{eq:Markovian-ME}) is thus only justified for $\lambda\ll\gamma$. Here we keep $\lambda/\gamma$ arbitrary as there are scenarios where Eq.~(\ref{eq:Markovian-ME}) also holds for strong measurements, see below.

%Here $\hat{A}\ket{a}=\xi_a\ket{a}$. With this approximation, $- \gamma \partial_D \mathcal{A}(D) \hat{\rho}_t(D) + \frac{\gamma^2}{8 \lambda} \partial_D^2 \hat{\rho}_t(D) = 0$. That is, a fast detector rapidly responds to changes in the system state and quickly reaches and remains in steady state with respect to the Fokker-Planck terms. Inserting the approximation for $\hat{\rho}_t(D)$ into Eq.~(\ref{eq:MAIN-RESULT}) results in $\mathcal{L}_{0} =  \int dD \mathcal{L}(D) \left[ \sum_{aa'} \pi_{aa'}(D) \mathcal{V}_{aa'} \right]$. The first order correction $\gamma^{-1}\mathcal{L}_{\rm corr}$ accounts for the lag of the detector due to its finite response time $\gamma^{-1}$, and as usual in linear response theory it can be written in terms of time-integrated correlation functions -- see the SM for details. We note that $\lambda$ plays a special role in the separation of timescales since it appears both in the first as well as the second line of Eq.~(\ref{eq:MAIN-RESULT}). In general, Eq.~(\ref{eq:Markovian-ME}) is thus only justified for $\lambda\ll\gamma$. Here we keep $\lambda/\gamma$ arbitrary as there are scenarios where Eq.~(\ref{eq:Markovian-ME}) also holds for strong measurements, see below.

We emphasize that Eq.~(\ref{eq:Markovian-ME}) describes arbitrary feedback protocols, both linear and nonlinear in $D$. As a consistency check, we recover the Wiseman-Milburn equation \cite{Wiseman-PRL-1993} from Eq.~(\ref{eq:MAIN-RESULT}) by employing the separation of timescales approximation to first order in $1/\gamma$, using a linear feedback Liouville superoperator $\mathcal{L}(D)\hat{\rho} = \mathcal{L}\hat{\rho} - i D[\hat{F},\hat{\rho}]$, with feedback Hamiltonian $\hat{F}$, and taking the infinite bandwidth limit (see SM). Our formalism thus generalizes the important earlier work of Ref.~\cite{Wiseman-PRL-1993} to nonlinear feedback protocols.

In the following, we highlight the usefulness of Eq.~(\ref{eq:MAIN-RESULT}) by studying protocols for power production in two toy models.

\begin{figure*}
\centering
\includegraphics[scale=0.7]{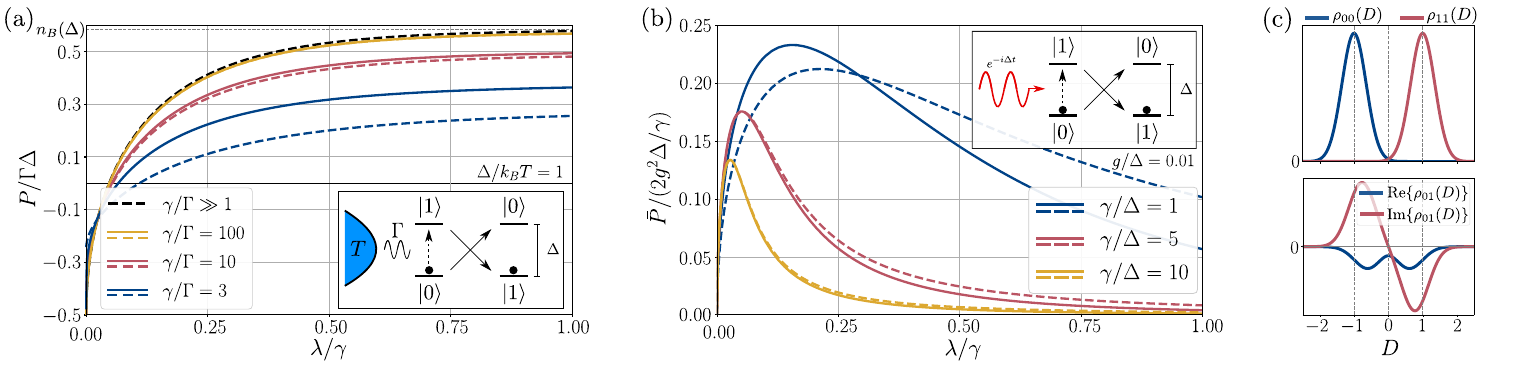}
\caption{Steady state power for classical (a) and quantum (b) toy models, varying the measurement strength $\lambda$. Solid lines obtained by numerically solving Eq.~(\ref{eq:MAIN-RESULT}), dashed lines obtained analytically using the separation of timescales technique. The separation of timescales assumption breaks down when system and detector timescales are comparable. (a) The inset illustrates a feedback protocol of a classical two-level system coupled to a thermal reservoir. When excited (dashed arrow), the levels are flipped (solid arrows), extracting energy. For strong measurements ($\lambda \gg \gamma$), the average occupation of the bath [$n_B(\Delta)$] sets an upper limit on extracted power, see dashed grey line, and is only reached for fast detectors ($\gamma / \Gamma \gg 1$) [cf.~Eq.~(\ref{eq:classical-power})]. For weak measurements $(\lambda \ll \gamma$), feedback is applied randomly and energy is dissipated into the reservoir. (b) The inset depicts a feedback protocol for a qubit, coherently driven by an external driving field. The protocol is identical to (a). For strong measurements, the power vanishes because of the quantum Zeno effect. For weak measurements, no power can be extracted as feedback is applied randomly. (c) Visualization of $\hat{\rho}_t(D)$ for the quantum toy model, with stationary matrix elements $\rho_{ab}(D) = \bra{a} \! \hat{\rho}_t(D) \! \ket{b}$. Here we use $g/\Delta = 0.01$ and $\gamma=\Delta=\lambda$. Top panel: diagonal elements of $\hat{\rho}_t(D)$. Bottom panel: real and imaginary part of $\rho_{01}(D)$.}
\label{fig:toy-panel}
\end{figure*}

\textit{Classical toy model.} By classical system, we refer to a situation with discrete energy levels, but where the density matrix remains diagonal in the energy basis at all times. This can be achieved either by suppressing quantum coherence by environmental noise, or by decoupling the diagonal and off-diagonal elements of $\hat\rho_t$ (see SM for details). Under these conditions, $[\hat{\rho}_t(D),\hat{A}]=0$ and the backaction term in Eq.~(\ref{eq:MAIN-RESULT}) has no influence on the dynamics. To facilitate a comparison between the classical and quantum models, we use the same notation. We consider a classical two-level system, with states $\ket{0}$ and $\ket{1}$, coupled to a thermal reservoir at temperature $T$, see inset of Fig.~\ref{fig:toy-panel}(a). The system and reservoir exchange energy quanta with energy $\Delta$ at rate $\Gamma$. The state of the system is continuously monitored by measuring the observable $\hat{A} = \hat{\sigma}_z$, with Pauli-Z operator $\hat{\sigma}_z = \ket{1}\!\bra{1} - \ket{0}\!\bra{0}$, such that whenever the measurement outcome $D<0$ ($D\geq0$) for an ideal detector (low noise and delay), the system resides in $\ket{0}$ ($\ket{1}$). Feedback is incorporated by flipping the levels according to the solid arrows in Fig.~\ref{fig:toy-panel}(a) when an excitation is detected, i.e., when $D$ changes sign, thereby extracting energy from the reservoir. The Hamiltonian is given by $\hat{H}(D) = [1-\theta(D)] \Delta \dyad{1} + \theta(D) \Delta \dyad{0}$, where $\theta(D)$ is the Heaviside step function. Note that $[\hat{H}(D),\hat{A}]=0$, ensuring that $\hat{\rho}_t(D)$ remains diagonal in the energy basis. The feedback protocol is represented by the Liouville superoperator
\begin{equation}
\mathcal{L}(D) = \left[ 1 - \theta(D) \right] \mathcal{L}_{-} + \theta(D) \mathcal{L}_{+},
\label{eq:Classical-Liouville-D}
\end{equation}
\noindent where  $\mathcal{L}_{-}\hat{\rho} = \Gamma n_B(\Delta) \mathcal{D}[\hat{\sigma}^\dagger]\hat{\rho} + \Gamma [n_B(\Delta)+1] \mathcal{D}[\hat{\sigma}] \hat{\rho} $ is the protocol applied for $D<0$, and $\mathcal{L}_{+}\hat{\rho} = \Gamma [n_B(\Delta)+1] \mathcal{D}[\hat{\sigma}^\dagger]\hat{\rho} + \Gamma n_B(\Delta) \mathcal{D}[\hat{\sigma}] \hat{\rho} $ is the protocol applied for $D\geq0$, with system ladder operator $\hat{\sigma} = \ket{0}\!\bra{1}$, and Bose-Einstein distribution $n_B(x) = [\exp(x/k_BT) - 1]^{-1}$, with $x$ denoting energy and $k_B$ the Boltzmann constant.

Employing the separation of timescales technique, using $\gamma \gg \Gamma$ with Eqs.~(\ref{eq:Markovian-ME}) and (\ref{eq:stationary-OU-distr}), the system evolves, to zeroth order in $1/\gamma$, according to the feedback Liouville superoperator
\begin{equation}
\begin{aligned}
\mathcal{L}_{0} = &\left[ (1-\eta) \mathcal{L}_{-} + \eta \mathcal{L}_{+} \right] \mathcal{V}_{00} \\
&+ \left[ \eta \mathcal{L}_{-} + (1-\eta) \mathcal{L}_{+} \right] \mathcal{V}_{11},
\end{aligned}
\label{eq:Classical-feedback-Liouvillian}
\end{equation}
\noindent where we introduced the feedback error probability $\eta = [1-\erf (2\sqrt{\lambda/\gamma})]/2$ for a single feedback event, where $\erf(\cdot)$ is the error function and $0 \leq \eta \leq 1/2$. Feedback is applied incorrectly when the measurement outcome does not reflect the true system state. Note that, weak (strong) measurements yield high (low) detector noise and increase (decrease) the error probability.

To zeroth order in $1/\gamma$, the average power production reads
\begin{equation}
P = \Gamma \Delta \Big[ (1-\eta) n_B(\Delta) - \eta [n_B(\Delta)+1] \Big],
\label{eq:classical-power}
\end{equation}
\noindent where $P > 0$ corresponds to extracting energy from the bath. For strong measurements ($\eta\rightarrow 0$), feedback is consistently applied correctly and energy is only extracted from the reservoir. The maximum extraction rate $P=\Gamma \Delta n_B(\Delta)$ is limited by the coupling $\Gamma$ and the average occupation $n_B(\Delta)$ of the bath. For weak measurements, feedback errors together with the asymmetry between excitation and de-excitation rates lead to a net dissipation of energy. Interestingly, the maximum dissipation rate $P=-\Gamma\Delta/2$ is independent of $n_B(\Delta)$. Equation (\ref{eq:classical-power}) is plotted with a black, dashed line in Fig.~\ref{fig:toy-panel}, illustrating the behavior for weak and strong measurements. Additionally, we computed the power by (i) numerically solving Eq.~(\ref{eq:MAIN-RESULT}) (solid colored lines), and (ii) using the separation of timescales technique to first order in $1/\gamma$ (dashed colored lines) (see SM for details). As $\gamma$ decreases, the extracted power decreases because the detector can no longer resolve fast changes in the system, missing opportunities to extract energy. The separation of timescales approximation gradually breaks down as $\gamma$ and $\Gamma$ become comparable.

Following Ref.~\cite{Esposito-EPL-2012}, in the long-time limit, Eq.~(\ref{eq:Classical-feedback-Liouvillian}) implies the detailed fluctuation theorem
\begin{equation}
\frac{P(-m)}{P(m)} = e^{m\left[ \Delta/k_BT - \ln \left( \frac{1-\eta}{\eta} \right) \right]}
\label{eq:fluctuation-thm}
\end{equation}

\noindent for the number of extracted energy quanta $m$, where $m>0$ ($m<0$) corresponds to extracting (dissipating) energy from the bath. The term $\Delta/T$ is the entropy change in the bath related to the exchange of a single quantum. The information term $\ln(\frac{1-\eta}{\eta})$ is given by the log-odds of not making an error and can be interpreted as the difference in information content between correctly and incorrectly applying feedback. Note that most information from the continuous measurement is discarded - it is only the information during a change in the system state that matters. In the error free limit, $\eta \rightarrow 0$, the information term diverges, illustrating absolute irreversibility, i.e., all excitations are extracted. See SM for a derivation of Eq.~(\ref{eq:fluctuation-thm}).

\textit{Quantum toy model.} We consider a qubit coherently driven by an external driving field, see inset of Fig.~\ref{fig:toy-panel}(b). Measurement and feedback are identical to the classical toy model, now extracting energy from the driving field. The feedback protocol is described by $\mathcal{L}_t(D)\hat{\rho} = -i[\hat{H}_t(D),\hat{\rho}]$ with Hamiltonian
\begin{equation}
\begin{aligned}
\hat{H}_t(D) = &[1-\theta(D)] \Delta \ket{1}\!\bra{1} \\ &+ \theta(D) \Delta \ket{0}\!\bra{0} + g\cos(\Delta t) \hat{\sigma}_x,
\end{aligned}
\label{eq:quantum-feedback-Hamiltonian}
\end{equation}
\noindent where $\Delta$ is the qubit level spacing, $g$ the strength of the qubit-driving field coupling, and $\hat{\sigma}_x$ the Pauli-X operator.

Separating system and detector timescales to first order in $1/\gamma$ results in system Liouville superoperator (details in SM)
\begin{equation}
\begin{aligned}
\left[ \mathcal{L}_{0} + \lambda \mathcal{D}[\hat{\sigma}_z] + \gamma^{-1}\mathcal{L}_{\text{corr}} \right] \! \hat{\rho} = -ig\cos(\Delta t)[\hat{\sigma}_x,\hat{\rho}] \\ + \tilde{\lambda} \mathcal{D}[\hat{\sigma}_z]\hat{\rho} - \frac{2\Delta g}{\gamma}D_0\cos(\Delta t)\hat{\sigma}_x,
\end{aligned}
\label{eq:quantum-feedback-Liouvillian}
\end{equation}
\noindent with effective dephasing rate $\tilde{\lambda} = \lambda + \Delta^2\ln(2)/2\gamma$, and coefficient $D_0 = 2 \sqrt{\lambda/\pi\gamma} \hspace{1mm} _{2}F_{2}(1/2,1/2;3/2,3/2;-4\lambda/\gamma)$, where $_{2}F_{2}(\cdot)$ is a generalized hypergeometric function. The first term on the RHS of Eq.~(\ref{eq:quantum-feedback-Liouvillian}) represents the coherent drive, while the second term describes dephasing due to measurement and feedback. The third term is a source for quantum coherence, stabilizing the coherence in the long-time limit. We emphasize that the first order correction is essential to compute the power as the steady state coherence vanishes to leading order, and hence, no power can be extracted. Note that the third term, which goes beyond leading order, can lead to negativities in $\hat{\rho}_t$, which is of no concern in the separation of timescales regime where the term is small. We stress that this term is trace preserving as $\hat{\sigma}_x$ is traceless.

The average power of the system is given by $P(t) = \trace\{[\partial_t \hat{H}(D)] \hat{\rho}_t\}$, where power is extracted (dissipated) when $P(t)>0$ ($P(t)<0$). Over one driving period $\tau=2\pi/\Delta$, the time averaged power reads
\begin{equation}
\bar{P} = \frac{2g^2\Delta}{\gamma} D_0 \frac{\Delta^2}{\Delta^2 + 4\tilde{\lambda}^2}.
\label{eq:quantum-power}
\end{equation}
\noindent For strong measurements $\lambda \gg \gamma$, the power vanishes because of the quantum Zeno effect. For weak measurements $\lambda \ll \gamma$, large detector noise leads to completely random feedback, and the power goes to zero because of the symmetric driving. This is highlighted in Fig.~\ref{fig:toy-panel}(b), where we plot Eq.~(\ref{eq:quantum-power}) as dashed lines. The solid lines were computed numerically by solving the full Eq.~(\ref{eq:MAIN-RESULT}). The corresponding steady state matrix elements of $\hat{\rho}_t(D)$ are visualized in Fig.~\ref{fig:toy-panel}(c) (details in SM). Similar to the classical toy model, the separation of timescales assumption breaks down when system and detector timescales are comparable.

\textit{Outline derivation main result.} To outline the main steps in the derivation of Eq.~(\ref{eq:MAIN-RESULT}), we start by describing the continuous measurement. For a single instantaneous measurement, the system state $\hat{\rho}_t$ transforms as
\begin{equation}
\hat{\rho}_t (z) = \hat{K}(z) \hat{\rho}_t \hat{K}^\dagger(z),
\end{equation}
\noindent where $\hat{K}(z)$ is the measurement operator for obtaining outcome $z$, obeying the completeness relation $\int dz \, \hat{K}^\dagger(z) \hat{K}(z) = \mathbb{1}$, $\trace\{ \hat{\rho}_t(z) \}$ is the probability of obtaining $z$, and $\int dz \, \hat{\rho}_t(z)$ is the system state for an unkown measurement outcome. Stressing that temporal coarse graining results in Gaussian noise for any measurement operator \cite{Jacobs-Contemp-phys-2006}, we consider Gaussian measurement operators \cite{Jacobs-Contemp-phys-2006,Bednorz-New-J-Phys-2012}
\begin{equation}
\hat{K}(z) = \left( \frac{2\lambda \delta t}{\pi} \right)^{1/4} e^{-\lambda \delta t \left( z - \hat{A} \right)^2},
\label{eq:measurement-operator}
\end{equation}
\noindent where $\delta t$ is the time between measurements. A weak continuous measurement is obtained by repeatedly measuring the system, taking the limit $\lambda \delta t \rightarrow 0$ for a fixed measurement strength $\lambda$. In this limit, the sequence of outcomes becomes a continuous signal $z(t)$.

The detector bandwidth $\gamma$ is introduced through a low-pass frequency filter \cite{Warszawski-JOpt-2002-I, Warszawski-JOpt-2002-II, Sarovar-PRA-2004, Sarovar-PRA-2005, Liu-PRA-2010, Wheatley-PRL-2010, Feng-PRA-2011}
\begin{equation}
D(t) = \int_{-\infty}^{t} ds \gamma e^{-\gamma(t-s)} z(s),
\label{eq:filtered-outcome}
\end{equation}
\noindent such that the measurement outcome $D(t)$ is a smoothened version of the signal $z(t)$. The filter reduces the high frequency measurement noise and introduces a detector delay. This provides a realistic detector model, but the filter is also necessary for nonlinear feedback protocols because higher orders of $z(t)$ are ill-defined due to its white noise spectrum which includes diverging frequencies \cite{Sarovar-PRA-2004,Sarovar-PRA-2005,Wheatley-PRL-2010}.

Feedback is incorporated by controlling the system time evolution in between measurements, i.e., making the Liouville superoperator $\mathcal{L}(D)$ dependent on the frequency filtered measurement outcome $D$. Combining time evolution due to measurements and due to the Liouvillian, we find Eq.~(\ref{eq:MAIN-RESULT}) in the continuous limit $\delta t \rightarrow 0$. The derivation can be carried out either in the framework of stochastic calculus following the methods outlined in Refs.~\cite{Jacobs-Contemp-phys-2006} and \cite{Wiseman/Milburn-book-2010}, or under the rules of conventional calculus. See details in SM.

\textit{Conclusions.} We have derived a Fokker-Planck master equation for continuous feedback control, describing the joint system-detector dynamics for detectors with finite bandwidth. By separating system and detector timescales, we obtain a Markovian master equation for the system alone, opening a new avenue for analytical modeling of nonlinear feedback protocols. The Markovian description further implies fluctuation theorems, providing insight into the connection between thermodynamics and information theory. With two simple toy models, we highlighted the usefulness of our formalism, showing that it can be applied to a large variety of systems in both the classical and quantum regimes. Future endeavors include extensions of the formalism to include non-Markovian effects and state-estimation feedback \cite{Belavkin-JMA-1992,Yanagisawa-PRL-2009}. %\cite{Debiossac-Nat.Com.-2020}.

\textit{Acknowledgments.} We thank Mark T.~Mitchison for fruitful discussions. This research was supported by grant number FQXi Grant Number: FQXi-IAF19-07 from the Foundational Questions Institute Fund, a donor advised fund of Silicon Valley Community Foundation. P.S. and B.A.A. were supported by the Swedish Research Council, grant number 2018-03921. P.P.P. acknowledges funding from the European Union's Horizon 2020 research and innovation programme under the Marie Sk\l{}odowska-Curie Grant Agreement No. 796700, from the Swedish Research Council (Starting Grant 2020-03362), and from the Swiss National Science Foundation (Eccellenza Professorial Fellowship PCEFP2\_194268).

\let\oldaddcontentsline\addcontentsline% Store \addcontentsline
\renewcommand{\addcontentsline}[3]{}% Make \addcontentsline a no-op
\bibliography{refdatabase,refdatabase2}
\let\addcontentsline\oldaddcontentsline% Restore \addcontentsline

\newpage
\widetext
\begin{center}
	\textbf{\large Supplemental Material: Quantum Fokker-Planck Master Equation for Continuous Feedback Control}
\end{center}
%%%%%%%%%% Merge with supplemental materials %%%%%%%%%%
%%%%%%%%%% Prefix a "S" to all equations, figures, tables and reset the counter %%%%%%%%%%
\setcounter{equation}{0}
\setcounter{figure}{0}
\setcounter{table}{0}
\setcounter{page}{1}
\makeatletter
\renewcommand{\theequation}{S\arabic{equation}}
\renewcommand{\thefigure}{S\arabic{figure}}

In this supplement, we provide detailed technical derivations for the results presented in the main text. Our main result, Eq.~(\ref{eq:MAIN-RESULT}) in the main text, is derived by two different methods in Sec.~\ref{sec:derivation}. Section \ref{sec:sep-of-TS} provides details on the separation of time-scales, which results in Eq.~(\ref{eq:Markovian-ME}) in the main text. Two different approaches are provided. Additional details on the numerical calculations are provided in Sec.~\ref{sec:general-numerics} and detailed calculations for the classical and quantum toy models are given in Secs.~\ref{sec:classical-toy} and \ref{sec:quantum-toy}, respectively. Equation and Figure numbers not preceded by an `S' refer to the main text.

\tableofcontents

\section{Derivations of the main result}
\label{sec:derivation}
\subsection{Conventional calculus}
\label{sec:derivation-I}

In this section, we derive the quantum Fokker-Planck master equation (QFPME) in Eq.~(\ref{eq:MAIN-RESULT}) in the main text by the means of conventional calculus. For compact notation, we introduce the measurement superoperator \cite{Jacobs-Contemp-phys-2006,Bednorz-New-J-Phys-2012}
\begin{equation}
\label{eq:mz}
\mathcal{M}(z) \hat{\rho} \equiv \hat{K}(z) \hat{\rho} \hat{K}^\dagger(z),\hspace{1.5cm}\hat{K}(z) = \left( \frac{2\lambda\delta t}{\pi} \right)^{1/4} e^{-\lambda\delta t(z-\hat{A})^2},
\end{equation}
\noindent where $z$ is the outcome, $\lambda$ measurement strength, and $\hat{A}$ the measured observable. To describe a continuous measurement, time is discretized into $n$ intervals $\delta t = (t-t_0)/n$, where $t_0$ and $t$ are the initial and final times, respectively. By successively applying time evolutions and measurements on the initial state $\hat{\rho}_{t_0}$, we get
\begin{equation}
\hat{\rho}_t(\{z_j\}_{j=1}^n) = \mathcal{M}(z_n) e^{\mathcal{L}(D_{n-1})\delta t} \cdots \mathcal{M}(z_2) e^{\mathcal{L}(D_{1})\delta t} \mathcal{M}(z_1) e^{\mathcal{L}\delta t} \hat{\rho}_{t_0},
\end{equation}
\noindent representing the joint state of the system and the sequence of outcomes $z_j$ obtained at times $t_j = t_0 + j\delta t$. Feedback is incorporated by the measurement dependent time evolutions in between measurements, where $D_j$ is the measurement outcome observed on the detector (which includes a low-pass filter with bandwidth $\gamma$) at time $t_j$. The relation between the filtered outcome $D_j$ and the unfiltered outcome $z_j$ is defined by 
\begin{equation}
\label{eq:dzrel}
D_j = \gamma \delta t \sum_{k=0}^j e^{-\gamma(j-k)\delta t}z_k,
\end{equation}
\noindent and is a discretized version of Eq.~(\ref{eq:filtered-outcome}). For a fixed measurement strength $\lambda$, we obtain a weak continuous measurement in the limit $\lambda \delta t \rightarrow 0$.

We may now write the joint state of the system and the measurement record $\{D_j\}_{j=1}^n$ as
\begin{equation}
\hat{\rho}_t(\{D_j\}_{j=1}^n) = \int dz_n \cdots dz_1 \prod_{r=0}^n \delta \left( D_r - \gamma \delta t \sum_{k=0}^r e^{-\gamma \delta t (r-k)} z_k \right) \hat{\rho}_t(\{z_j\}_{j=1}^n),
\label{eq:path-integral}
\end{equation}
\noindent with $z_0$ specifying the initial value of $z(t)$. Equation (\ref{eq:path-integral}) results in
\begin{equation}
\hat{\rho}_t(\{D_j\}_{j=1}^n) = M(D_n|D_{n-1}) e^{\mathcal{L}(D_{n-1})\delta t} \cdots e^{\mathcal{L}(D_{1})\delta t} M(D_1|D_{0}) e^{\mathcal{L}\delta t} \hat{\rho}_{t_0},
\end{equation}
\noindent with the new measurement operator $M(D|D') = \mathcal{M}\left(\frac{D-D'e^{-\gamma\delta t}}{\gamma \delta t}\right)/\gamma\delta t$ for the measurement outcome $D$ given that the outcome in the previous timestep was $D'$. Using that $\hat{\rho}_t(D_n) = \int dD_{n-1} \cdots dD_1 \hat{\rho}_t(\{D_j\}_{j=1}^n)$ leads to
\begin{equation}
\hat{\rho}_{t+\delta t}(D) = \int dD' M(D|D') e^{\mathcal{L}(D')\delta t} \hat{\rho}_t(D'),
\label{eq:update-rule}
\end{equation}
%´
\noindent where we substituted $D_{n-1} \rightarrow D'$ and $D_n \rightarrow D$. Finally, to first order in $\delta t$, the measurement operator $M(D|D')$ reads
\begin{equation}
M(D|D') \hat{\rho} \approx \delta(D-D')\hat{\rho} + \delta t \left[ \lambda \delta(D-D') \mathcal{D}[\hat{A}]\hat{\rho} - \gamma \delta'(D-D')\mathcal{A}(D)\hat{\rho} + \frac{\gamma^2}{8\lambda}\delta''(D-D')\hat{\rho} \right],
\label{eq:M(D|D')}
\end{equation}
\noindent where $\delta'(D-D')$ and $\delta''(D-D')$ denote the first and second derivative with respect to $D$ on the Dirac delta function. Equation (\ref{eq:M(D|D')}) was found by using $\hat{A}\ket{a}=\xi_a\ket{a}$ and the inverse Fourier transform
\begin{equation}
\sqrt{\frac{2\lambda}{\pi\gamma^2\delta t}} e^{-2\lambda\delta t\left(\frac{D-D'e^{-\gamma\delta t}}{\gamma \delta t}-\frac{\xi_a+\xi_{a'}}{2}\right)^2} \approx \frac{1}{2\pi} \int_{-\infty}^\infty d\omega e^{-i\omega(D-D')} e^{\frac{1}{8}\gamma\omega\delta t\left[4i(\xi_a+\xi_{a'}) - \frac{\gamma}{\lambda}\omega - 8iD' \right]}, 
\end{equation}
\noindent where we used $e^{-\gamma\delta t}\approx 1-\gamma\delta t$. The first order expansion of the LHS is found by expanding the second exponential under the integral, and then computing the integral. Inserting Eq.~(\ref{eq:M(D|D')}) in Eq.~(\ref{eq:update-rule}), letting $\delta t \rightarrow dt$ results in Eq.~(\ref{eq:MAIN-RESULT}).

Finally, we emphasize that Eq.~(\ref{eq:MAIN-RESULT}) in the main text preserves the trace of $\hat{\rho}_t = \int dD \hat{\rho}_t(D)$. To describe a normalized probability distribution over $D$, both $\hat{\rho}_t(D)$ and $\partial_D\hat{\rho}_t(D)$ must vanish as $|D|\rightarrow \infty$. Thus, by integrating Eq.~(\ref{eq:MAIN-RESULT}) over $D$, the last two terms vanish. The remaining two terms, i.e., $\mathcal{L}(D)$ and $\mathcal{D}[\hat{A}]$, are trace preserving, implying that Eq.~(\ref{eq:MAIN-RESULT}) is trace preserving as well. It follows that $\trace\{\hat{\rho}_t\}=1$, and that $\int dD P_t(D)=1$, where $P_t(D)=\trace\{\hat{\rho}_t(D)\}$.

\subsection{Stochastic calculus}
\label{sec:derivation-II}
Equation (\ref{eq:MAIN-RESULT}) in the main text may also be derived using the tools of It\^o stochastic calculus. To this end, we consider the conditional density matrix which changes over time as \cite{wiseman:book}
\begin{equation}
\label{eq:updatec}
\hat{\rho}_c(t+dt) = e^{\mathcal{L}(D)dt}\frac{\mathcal{M}(z)\hat{\rho}_c(t)}{{\rm Tr}\{\mathcal{M}(z)\hat{\rho}_c(t)\}},
\end{equation}
where $z$ denotes the (unfiltered) measurement outcome at time $t$ and $\mathcal{M}(z)$ is given in Eq.~\eqref{eq:mz}. Note that due to the denominator on the right-hand side, this equation is nonlinear. We now introduce the Wiener increment 
\begin{equation}
\label{eq:wiener}
dW = 2\sqrt{\lambda}d t\left(z-\langle \hat{A}\rangle_c\right),\hspace{2cm}\langle\hat{A}\rangle_c = {\rm Tr}\{\hat{A}\hat{\rho}_c(t)\},
\end{equation}
which has zero mean and obeys $dW^2=dt$ \cite{Jacobs-Contemp-phys-2006}. With Eq.~\eqref{eq:wiener}, we may eliminate $z$ from Eq.~\eqref{eq:updatec}. Expanding Eq.~\eqref{eq:updatec} to first order in $dt$ (second order in $dW$) then results in the Belavkin equation \cite{belavkin:1989}
\begin{equation}
\label{eq:belavkin}
d\hat{\rho}_c\equiv \hat{\rho}_c(t+dt)-\hat{\rho}_c(t) = \mathcal{L}(D) dt\hat{\rho}_c+\lambda dt \mathcal{D}[\hat{A}]\hat{\rho}_c+\sqrt{\lambda} dW \{\hat{A}-\langle \hat{A}\rangle_c,\hat{\rho}_c\},
\end{equation}
with $\mathcal{D}[\hat{a}]\hat{\rho} = \hat{a}\hat{\rho}\hat{a}^\dagger-\frac{1}{2}\{\hat{a}^\dagger\hat{a},\hat{\rho}_c\}$. From Eq.~\eqref{eq:dzrel}, we find the stochastic differential equation 
\begin{equation}
\label{eq:stochD}
dD=\gamma (z-D)dt = \gamma (\langle A\rangle_c-D)dt+\frac{\gamma}{2\sqrt{\lambda}}dW.
\end{equation}
Finally, we note that the quantity of interest can be written as
\begin{equation}
\label{eq:densDstoch}
\hat{\rho}_t(D_0) = E[\hat{\rho}_c\delta (D-D_0)],\hspace{2cm} \partial_t\hat{\rho}_t(D_0) = \frac{E[d(\hat{\rho}_c\delta (D-D_0))]}{dt},
\end{equation}
where $E[\cdot]$ denotes the  average over the full history of measurement outcomes. The Dirac delta in the average ensures that we preserve the current (last) measurement outcome. Taking the trace over $\hat{\rho}_t(D_0)$, we obtain  the probability of observing $D$ at time $t$, $P_t(D_0)=E[\delta(D-D_0)]$ while integrating over $D_0$ provides the normalized, unconditional density matrix $\hat{\rho}_t = E[\hat{\rho}_c]$. The stochastic product and chain rules imply
\begin{equation}
\label{eq:chain}
d(\hat{\rho}_c\delta (D-D_0))=(d\hat{\rho}_c)\delta (D-D_0)+\hat{\rho}_cd\delta (D-D_0)+(d\hat{\rho}_c)d\delta (D-D_0),
\end{equation}
and
\begin{equation}
\label{eq:diffdelta}
d\delta (D-D_0) = \delta'(D-D_0)\left[\gamma(\langle A\rangle_c-D_0)dt+\frac{\gamma}{2\sqrt{\lambda}}dW\right]+\frac{1}{2}\delta''(D-D_0)\frac{\gamma^2}{4\lambda} dt,
\end{equation}
Inserting Eq.~\eqref{eq:chain} into Eq.~\eqref{eq:densDstoch}, we recover Eq.~(\ref{eq:MAIN-RESULT}) in the main text with the help of Eqs.~(\ref{eq:diffdelta},\ref{eq:belavkin}) and by employing
\begin{equation}
\label{eq:ddelprob}
{\rm E}[\delta'(D-D_0)\hat{\rho}_c] = -\partial_{D_0} \hat{\rho}_t(D_0),\hspace{2cm}{\rm E}[\delta''(D-D_0)\hat{\rho}_c] = \partial^2_{D_0} \hat{\rho}_t(D_0).
\end{equation}

\section{Separation of timescales}
\label{sec:sep-of-TS}
In this section, we provide details on the treatment of the regime where the detector and the system are governed by different time-scales. This results in Eq.~(\ref{eq:Markovian-ME}) in the main text. We provide two different approaches, one based on Nakajima-Zwanzig projection operators, Sec.~\ref{sec:sep-of-TS-Patrick}, and one based on multiple time scale perturbation theory, Sec.~\ref{sec:multitime}.

\subsection{Nakajima-Zwanzig approach}
\label{sec:sep-of-TS-Patrick}

To employ Nakajima-Zwanzig projection operators \cite{nakajima:1958,zwanzig:1960}, we first re-write Eq.~(\ref{eq:MAIN-RESULT}) in the main text as
\begin{equation}
\label{eq:masternz}
\begin{aligned}
&\partial_t \hat{\rho}_t(D) = \mathcal{L}_\lambda(D)\hat{\rho}_t(D)+\mathcal{F}(D)\hat{\rho}_t(D),\\
\mathcal{L}_\lambda(D) = \mathcal{L}(D)&  + \lambda \mathcal{D}[\hat{A}],\hspace{2cm}\mathcal{F}(D)=- \gamma \partial_D \mathcal{A}(D) + \frac{\gamma^2}{8 \lambda} \partial_D^2,
\end{aligned}
\end{equation}
where the individual terms are defined in the main text. We now introduce the projection superoperator
\begin{equation}
\label{eq:nzp}
\mathcal{P} \hat{\rho}_t(D) = \sum_{aa'}\pi_{aa'}(D)\mathcal{V}_{aa'}\int d D'\hat{\rho}_t(D'),\hspace{1.5cm}\mathcal{Q}=1-\mathcal{P},
\end{equation}
where $\pi_{aa'}(D)$ is given in Eq.~(\ref{eq:stationary-OU-distr})
and $\mathcal{V}_{aa'}\hat{\rho}=|a\rangle\langle a|\hat{\rho}|a'\rangle\langle a'|$. We note that we have $\mathcal{P}\mathcal{F}(D) = \mathcal{F}(D)\mathcal{P} = 0$. Using these superoperators, we can show that
\begin{equation}
\label{eq:nomarknz}
\partial_t \mathcal{P} \hat{\rho}_t(D)  =\mathcal{P}\mathcal{L}_\lambda(D) \mathcal{P}\hat{\rho}_t(D)+\mathcal{P}\mathcal{L}_\lambda(D)\mathcal{Q} \int_{0}^{t}dse^{\mathcal{Q}[\mathcal{L}_\lambda(D)+\mathcal{F}(D)](t-s)}\mathcal{Q}\mathcal{L}_\lambda(D) \mathcal{P}\hat{\rho}_s(D),
\end{equation}
where we assumed that the initial state fulfills $\mathcal{Q} \hat{\rho}_0(D)=0$.

We now approximate the last equation assuming a separation of time-scales. For bookkeeping purposes, we assume $\mathcal{L}_\lambda(D)\propto \Gamma$ and $\mathcal{F}(D)\propto \gamma$, with $\gamma\gg\Gamma$. To lowest order in $\Gamma$, we may then drop $\mathcal{L}_\lambda(D)$ in the exponential in Eq.~\eqref{eq:nomarknz}. Furthermore, we make a Markov approximation, replacing the time argument of the density matrix under the integral $\hat{\rho}_s(D)\rightarrow \hat{\rho}_t(D)$ and we extend the integral to minus infinity. This is justified when the integrand vanishes much faster than the time-scale over which $\hat{\rho}_t$ changes. This results in the differential equation
\begin{equation}
\label{eq:septime2}
\partial_t \mathcal{P} \hat{\rho}_t(D) =\mathcal{P}\mathcal{L}_\lambda(D) \mathcal{P}\hat{\rho}_t(D)-\mathcal{P}\mathcal{L}_\lambda(D)\mathcal{F}_{\rm d}^{-1}(D)\mathcal{L}_\lambda(D) \mathcal{P}\hat{\rho}_t(D),
\end{equation}
where we introduce the Drazin inverse \cite{mandal:2016,scandi:2019}
\begin{equation}
\label{eq:drazin}
\mathcal{F}_{\rm d}^{-1}(D) = -\int_{0}^{\infty}dt e^{\mathcal{F}(D)t}\mathcal{Q}.
\end{equation}
Integrating Eq.~\eqref{eq:septime2} over $D$ reproduces Eq.~(\ref{eq:Markovian-ME}) in the main text with
\begin{equation}
\label{eq:lcorr}
\mathcal{L}_0 =\int dD\mathcal{L}(D)\sum_{aa'}\pi_{aa'}(D)\mathcal{V}_{aa'} ,\hspace{1cm}
\mathcal{L}_{\rm corr} =  -\gamma\int dD\mathcal{L}_\lambda(D)\mathcal{Q}\mathcal{F}_{\rm d}^{-1}(D)\mathcal{L}_\lambda(D)\sum_{aa'}\pi_{aa'}(D)\mathcal{V}_{aa'}.
\end{equation}

We now introduce the generalized Hermite polynomials of variance $\sigma=\gamma/(8\lambda)$
\begin{equation}
\label{eq:hermitesigm}
He_n^{[\sigma]} \equiv \left( \frac{\sigma}{2} \right)^{n/2} H_n\left( \frac{x}{\sqrt{2\sigma}} \right),
\end{equation}
where $H_n(x)=(-1)^ne^{x^2}\partial_x^n e^{-x^2}$ are the standard physicist's Hermite polynomials. The generalized Hermite polynomials fulfill the orthogonality condition
\begin{equation}
\label{eq:hermorth}
\int_{-\infty}^\infty dx He_n^{\lfloor \sigma\rfloor}(x)He_m^{\lfloor \sigma\rfloor}(x)\frac{e^{-\frac{x^2}{2\sigma}}}{\sqrt{2\pi\sigma}} = \delta_{n,m}n!\sigma^n.
\end{equation}
With these definitions, it can be shown that (for $n>0$)
\begin{equation}
\label{eq:eigenfokk}
\int_{-\infty}^\infty dD He_n^{\lfloor \sigma\rfloor}\left(D-\frac{\xi_a+\xi_{a'}}{2}\right)\langle a|[\mathcal{F}^{-1}_{\rm d}(D)|a\rangle\langle a'|]|a'\rangle g(D) = -\frac{1}{n\gamma}\int_{-\infty}^\infty dD He_n^{\lfloor \sigma\rfloor}\left(D-\frac{\xi_a+\xi_{a'}}{2}\right)g(D),
\end{equation}
for any function $g(D)$. For $n=0$, the integral on the left-hand side vanishes. The generalized Hermite polynomials can thus be understood as the left eigenvectors of $\mathcal{F}(D)$ and $\mathcal{F}^{-1}_{\rm d}(D)$ with eigenvalues $-n\gamma$ and $-1/(n\gamma)$ respectively (for $n>0$). We further expand the matrix elements of the Liouvillian
\begin{equation}
\label{eq:lioherm}
\mathcal{L}_{bb',aa'}^n =\frac{1}{\sqrt{n!\sigma^{n}}}\int_{-\infty}^{\infty} \frac{dD}{\sqrt{2\pi\sigma}}\langle b|[\mathcal{L}(D)|a\rangle\langle a'|]|b'\rangle He_n^{\lfloor \sigma\rfloor}\left(D-\frac{\xi_a+\xi_{a'}}{2}\right)e^{-\frac{1}{2\sigma}\left(D-\frac{\xi_a+\xi_{a'}}{2}\right)^2}.
\end{equation}
We note that for $n=0$, the last expression reduces to the matrix elements of $\mathcal{L}_0$, cf.~Eq.~\eqref{eq:lcorr}.
With the help of Eqs.~(\ref{eq:eigenfokk},\ref{eq:lioherm}), we can write the matrix elements of $\mathcal{L}_{\rm corr}$ in Eq.~\eqref{eq:lcorr} as
\begin{equation}
\label{eq:lioucorrmat}
\mathcal{L}^{\rm corr}_{bb',aa'} = \langle b|[\mathcal{L}_{\rm corr}|a\rangle\langle a'|]|b'\rangle= \sum_{n=1}^{\infty}\frac{1}{n}\sum_{k=0}^n\frac{\sqrt{n!/k!}}{(n-k)!}\sum_{cc'}\frac{(\xi_a+\xi_{a'}-\xi_c-\xi_{c'})^{n-k}}{(2\sqrt{\sigma})^{n-k}}\mathcal{L}_{bb',cc'}^n\mathcal{L}_{cc',aa'}^k.
\end{equation}

\subsubsection{Linear feedback}
Here we consider a feedback Liouvillian of the form
\begin{equation}
\label{eq:linearfb}
\mathcal{L}_\lambda(D)\hat{\rho} = \mathcal{L}\hat{\rho}+\lambda\mathcal{D}[\hat{A}]\hat{\rho}-iD[\hat{F},\hat{\rho}].
\end{equation}
In the inifinite bandwidth limit, this scenario is described by the Markovian master equation derived by Wiseman and Milburn \cite{Wiseman-PRL-1993}.
For the zeroth order, we find
\begin{equation}
\label{eq:lio0linfb}
\mathcal{L}_0\hat{\rho}=\mathcal{L}\hat{\rho}-i[\hat{F},\mathcal{A}\hat{\rho}],\hspace{2cm}\mathcal{A}\hat{\rho}=\frac{1}{2}\{\hat{A},\hat{\rho}\}.
\end{equation}
The first order reduces to
\begin{equation}
\label{eq:masterlin2}
\gamma^{-1}\mathcal{L}_{\rm corr}\hat{\rho}= \frac{1}{4\lambda}\mathcal{D}[\hat{F}]\hat{\rho}-\frac{i}{\gamma}[\hat{F},[\mathcal{L},\mathcal{A}]\hat{\rho}]-\frac{1}{2\gamma}\Big[\hat{F},\big[[\hat{F},\hat{A}],\mathcal{A}\hat{\rho}\big]\Big].
\end{equation}
In the limit $\gamma\rightarrow \infty$, we recover the master equation by Wiseman and Milburn \cite{Wiseman-PRL-1993}. The last two terms constitute finite bandwidth correction terms to this well-known equation. We note that the first term in $\mathcal{L}_{\rm corr}$ is linear in $\gamma$ and thus also contributes to the infinite bandwidth limit. The reason for this is that in this limit, $D$ itself, and thus the eigenvalues of $\mathcal{L}_{\lambda}(D)$, may become very large. In particular, the standard deviation of $D$ scales as $\sqrt{\gamma}$ resulting in an extra factor of $\gamma$ on the right-hand side of Eq.~\eqref{eq:lcorr}.

\subsubsection{Threshold feedback}
\label{sec:threshfeed}
In the main text, we focus on feedback Liouvillians of the form
\begin{equation}
\label{eq:threshfb}
\mathcal{L}_\lambda(D) = \lambda\mathcal{D}[\hat{A}]+\theta(D)\mathcal{L}_++[1-\theta(D)]\mathcal{L}_-,
\end{equation}
where $\theta(x)=1$ for $x\geq0$ denotes the Heaviside theta function.
For the zeroth order, we find
\begin{equation}
\label{eq:lio0threshfb}
\mathcal{L}_0=\mathcal{L}_+\mathcal{E}_++\mathcal{L}_-\mathcal{E}_-,
\end{equation}
where we introduced the superoperators
\begin{equation}
\label{eq:supes}
\mathcal{E}_{\pm} = \sum_{aa'} \mathcal{V}_{aa',aa'}\frac{1}{2}\left(1\pm{\rm erf}\left[\sqrt{\lambda/\gamma}(\xi_a+\xi_{a'})\right]\right).
\end{equation}
For the $n\neq0$ coefficients, we find
\begin{equation}
\label{eq:lionthreshfb}
\mathcal{L}^n_{bb',aa'} = \left(\mathcal{L}_+-\mathcal{L}_-\right)_{bb',aa'}\frac{(-1)^{n-1}}{\sqrt{n!2^n\pi}}e^{-\frac{\lambda}{\gamma}(\xi_a+\xi_{a'})^2}H_{n-1}\left[\sqrt{\lambda/\gamma}(\xi_a+\xi_{a'})\right].
\end{equation}

\subsection{Multiple time scale perturbation approach in Fock-Liouville space}
\label{sec:multitime}

Here we present an alternative approach, using multiple time scale perturbation theory \cite{strogatz2018nonlinear}, for deriving the reduced master equation (\ref{eq:Markovian-ME}) of the main text.
We analyze the problem in Fock-Liouville space, where density matrices are converted to column vectors and superoperators become matrices \cite{Gyamfi_2020, Manzano2020}. 
When we apply this formalism to a classical model in Sec.~\ref{sec:classical-model-rateEq}, the subspaces corresponding to the classical populations and quantum coherences decouple, leading to classical rate equations.
%The multiple time scale approach captures both the slow- and the (transient) fast-scale dynamics, and provides a master equation for the density matrix when the detector degree of freedom is integrated out.

\subsubsection{Extension to Fock-Liouville space}
Equation (\ref{eq:MAIN-RESULT}) of the main text can be rewritten by introducing a vectorized form of the density matrix $|\rho \rangle \rangle$, containing the entries of the original matrix stacked in a single vector with $N^2$ elements, where $N$ is the dimension of the Hilbert space~\cite{Gyamfi_2020, Manzano2020}. The master equation (\ref{eq:MAIN-RESULT}) then becomes
\begin{equation}
    \label{eq:linear_ME}
    \partial_t |\rho_t(D)\rangle \rangle = \mathcal{L}_D |\rho_t(D) \rangle \rangle + \lambda \mathcal{D}_A |\rho_t(D)\rangle \rangle + \gamma \mathcal{F}|\rho_t(D) \rangle \rangle .
\end{equation}Here $\mathcal{L}_D$ and $\mathcal{D}_A$ are $N^2\times N^2$ matrices with entries
\begin{eqnarray}
    \langle \langle bb' | \mathcal{L}_D | {aa'}\rangle\rangle &=& (\mathcal{L}_D)_{bb',aa'} = \langle b | \mathcal{L}(D) \left[ | a \rangle \langle a' |\right] | b' \rangle,
    \\
    \label{eq:DA}
    \langle \langle bb'| \mathcal{D}_A | aa'\rangle\rangle &=& \left(\mathcal{D}_A\right)_{bb',aa'} = -\frac{1}{2}\left( \xi_a - \xi_{a'} \right)^2 \delta_{ab}\delta_{a'b'},
\end{eqnarray}
where $\hat{A}|a\rangle = \xi_a | a \rangle$ as in the main text.
The elements of $\mathcal{L}_D$ are functions of $D$, while the elements of the diagonal matrix $\mathcal{D}_A$ are numbers.
The operator $\mathcal{F}$ is a diagonal matrix of operators, whose elements $\mathcal{F}_{aa',aa'} \equiv \mathcal{F}_{aa'}$ are Ornstein–Uhlenbeck (OU) operators in $D$ space
\begin{equation}
\label{eq:ou}
\mathcal{F}_{aa'}\phi = \partial_D \left[ D - \frac{\left( \xi_a + \xi_{a'} \right)}{2} \right]\phi + \frac{\gamma}{8\lambda} \partial_D^2\phi,
\end{equation}
with Gaussian stationary distributions centered at $(\xi_a+\xi_{a'})/2$ with variance $\gamma/8\lambda$, i.e., $\mathcal{F}_{aa'}\pi_{aa'}(D)=0$ with $\pi_{aa'}$ given in Eq.~(\ref{eq:stationary-OU-distr}). Note that in contrast to the Sec.~\ref{sec:sep-of-TS-Patrick}, here we redefined OU operators in a dimensionless form (we factor out $\gamma$ from Eq.~(\ref{eq:masternz}) to get Eq.~(\ref{eq:ou})) for convenience of applying the perturbation scheme as shown in the next subsection. However, to avoid clutter we are not introducing any new notation for the OU operators in this section and using $\mathcal{F}$ to refer to its dimensionless form as shown in Eq.~(\ref{eq:ou}).

The matrices $\mathcal{L}_D$ and $\mathcal{D}_A$ are the Fock-Liouville representations of the superoperators $\mathcal{L}(D)$ and $\mathcal D[\hat A]$ appearing in the first line of Eq.~(\ref{eq:MAIN-RESULT}), and $\mathcal{F}$ is equivalent to the superoperator appearing on the second line of that equation.

We also define the $N^2\times N^2$ diagonal matrix
\begin{equation}\label{eq:Gdef}
    G(D)=\sum_{aa'} |aa'\rangle\rangle\pi_{aa'}(D)\langle\langle aa'|,
\end{equation}
such the $aa'$ diagonal element of $G(D)$ is the stationary distribution of $\mathcal{F}_{aa'}$.
The matrix $G(D)$ is the Fock-Liouville counterpart of the superoperator $\sum_{aa^\prime}\pi_{aa^\prime}(D)\mathcal{V}_{aa^\prime}$ appearing after Eq.~(\ref{eq:Markovian-ME}) of the main text.

\subsubsection{Perturbation scheme}
To derive Eq.~(\ref{eq:Markovian-ME}) of the main text from Eq.~(\ref{eq:MAIN-RESULT}) using the multiple time scale method, we start with Eq.~(\ref{eq:linear_ME}) and define two time scales of the system: a slow time scale $\tau_1=t$ and a fast time scale $\tau_2=\gamma t /\Gamma $, where $\Gamma$ is defined in the main text. The smallness parameter for the method is  $\epsilon=\Gamma/\gamma$. We now extend $| \rho_t(D)\rangle\rangle$ to its \textit{two-timed} \cite{strogatz2018nonlinear} analogue $|\rho(D,\tau_1,\tau_2)\rangle\rangle$
and rewrite Eq.~(\ref{eq:linear_ME}) as 
\begin{equation} \label{two_timed_eqn_full}
    \epsilon \left[\partial_{\tau_1}-\mathcal{L}_D-\lambda\mathcal{D}_A \right]|\rho(D,\tau_1,\tau_2)\rangle\rangle = -\left[ \partial_{\tau_2} -\Gamma\mathcal{F}\right]|\rho(D,\tau_1,\tau_2)\rangle\rangle.
\end{equation}
Next, we expand our two-timed state vector $|\rho(D,\tau_1,\tau_2)\rangle\rangle$ in a series,
\begin{equation}\label{two_timed_expansion}
    |\rho(D,\tau_1,\tau_2)\rangle\rangle=\sum_{k=0}^{\infty}\epsilon^k |\rho^{[k]}(D,\tau_1,\tau_2)\rangle\rangle.
\end{equation}
Substituting Eq.~\eqref{two_timed_expansion} into Eq.~(\ref{two_timed_eqn_full}) and collecting terms by orders
of $\epsilon$ we obtain the set of equations:
\begin{eqnarray}
\label{0th_order}
     \left[ \partial_{\tau_2} -\Gamma\mathcal{F}\right]|\rho^{[0]}(D,\tau_1,\tau_2)\rangle\rangle &=& \, 0,
     \\
\label{kth_order}
         \left[ \partial_{\tau_2} -\Gamma\mathcal{F}\right]|\rho^{[k]}(D,\tau_1,\tau_2)\rangle\rangle &=& -\left[\partial_{\tau_1}-\mathcal{L}_D-\lambda\mathcal{D}_A \right]|\rho^{[k-1]}(D,\tau_1,\tau_2)\rangle\rangle
         ,\quad k \ge 1.
\end{eqnarray}

The $0$th order Eq.~\eqref{0th_order} implies that $|\rho^{[0]}(D,\tau_1,\tau_2)\rangle \rangle$ evolves under the OU operator $\mathcal{F}$ on the fast time scale. The general solution to this equation can be written as
\begin{equation}\label{0th_sol}
    |\rho^{[0]}(D,\tau_1,\tau_2)\rangle\rangle=e^{\tau_2 \Gamma\mathcal{F}}|\rho^{[0]}(D,\tau_1,0)\rangle\rangle.
\end{equation}
The general solution to the $k$th order Eq.~\eqref{kth_order} is
\begin{equation}\label{kth_sol}
       |\rho^{[k]}(D,\tau_1,\tau_2)\rangle\rangle=e^{\tau_2 \Gamma\mathcal{F}}|\rho^{[k]}(D,\tau_1,0)\rangle\rangle  - \int_0^{\tau_2} \dd s \  e^{(\tau_2 - s)\Gamma\mathcal{F}}\left[ \partial_{\tau_1} -\mathcal{L}_D- \lambda\mathcal{D}_A)\right]|\rho^{[k-1]}(D,\tau_1,s)\rangle\rangle.
\end{equation}
The second term on the right side of Eq.~(\ref{kth_sol}) may lead to  secular terms, that is terms that grow linearly with $\tau_2$.
For consistency with the perturbation scheme, such terms must be removed \cite{strogatz2018nonlinear}.
%\textcolor{red}{[MTS reference here!]} 
Secular terms arise in Eq.~(\ref{kth_sol}) if the source term on the right side of Eq.~(\ref{kth_order}) contains a component inside the nullspace of the operator $\mathcal{F}$.
We therefore remove secular terms by imposing the condition that this source term contains no component in the nullspace of $\mathcal{F}$. This implies
\begin{equation}\label{secular_condition}
    \int_{-\infty}^{\infty} \dd D \left[\partial_{\tau_1}-\mathcal{L}_D-\lambda\mathcal{D}_A \right]|\rho^{[k-1]}(D,\tau_1,\tau_2)\rangle\rangle =0.
\end{equation}
Notice that, in removing the secular terms at $k$th order, Eq.~(\ref{secular_condition}) imposes a condition on the $(k-1)$th
order solution. 
Thus, to completely specify a solution to any order of the perturbation scheme we need to impose the condition that the source term in the next order of the perturbation equation exists outside the nullspace of the operator $\mathcal{F}$. Once this perturbative solution for the two-timed state vector $ |\rho^{[k]}(D,\tau_1,\tau_2)\rangle\rangle$ is obtained, reverting back to the original state vector $|\rho_t^{[k]}(D)\rangle\rangle$ leads to the multiple time scale solution to the problem.

Solutions obtained from the multiple time-scale approach capture
the dynamics both in the fast and slow time-scales but often we
are more interested in the slow time-scale dynamics where we
neglect the effect of the fast or transient dynamics. To obtain the slow time scale dynamics after the transient time, we set $\tau_2 \to \infty$ in Eq.~\eqref{0th_sol} and Eq. \eqref{kth_sol} and then substitute back $\tau_1=t$. Under this separation of time scale assumption the $0$th order system-detector distribution is given as
\begin{equation}\label{0th_sep_time}
     |\rho^{[0]}_{t}(D)\rangle\rangle=G(D)|\bar{\rho}^{[0]}_{t}\rangle\rangle,
\end{equation}
where
$|\bar{\rho}^{[0]}_{t}\rangle\rangle$ is yet to be determined. Similarly for $k\ge 1$ we have
\begin{equation}\label{kth_sep_time}
    |\rho^{[k]}_{t}(D)\rangle\rangle =G(D)|\bar{\rho}^{[k]}_{t}\rangle\rangle +\Gamma^{-1}\mathcal{F}^{+} \left[\partial_t -\mathcal{L}_D -\lambda\mathcal{D}_A\right]|\rho^{[k-1]}_{t}(D)\rangle\rangle,
\end{equation}
with $|\bar{\rho}^{[k]}_{t}\rangle\rangle$ yet to be determined. $\mathcal{F}^{+}$ is a diagonal matrix containing pseudo-inverses of the OU operators in the corresponding diagonal elements of $\mathcal{F}$ matrix
\begin{equation}
    \mathcal{F}^{+}=-\int_0^{\infty} \dd z \  e^{z\mathcal{F}} \ (I-\mathcal{P}_0),
\end{equation}
where $I$ is the identity matrix in the Fock-Liouville space and $\mathcal{P}_0$ is the null-space projection operator defined as
\begin{equation}
    \mathcal{P}_0|f(D)\rangle\rangle=G(D)\left (\int_{-\infty}^{\infty}\dd D |f(D)\rangle\rangle \right).
\end{equation}
This operator $\mathcal{F^{+}}$ is equivalent to the Drazin inverse defined in Eq.~\eqref{eq:drazin} but now it is considered in the Fock-Liouville space. These coefficient vectors $|\bar{\rho}^{[j]}_{t}\rangle\rangle$ with $j\ge0$ can be understood as the marginalized density matrix elements written as vectors of the system when the detector variable $D$ has been integrated out
\begin{equation}
\label{eq:coeff_vect}
    |\bar{\rho}^{[j]}_{t}\rangle\rangle=\int_{-\infty}^{\infty} \dd D |\rho^{[j]}_{t}(D)\rangle\rangle.
\end{equation}
To determine these vectors, we use the secularity removal condition, Eq.~\eqref{secular_condition}, written in the original variable $t$
\begin{equation}\label{secular_condn_sep_time}
    \partial_t |\bar{\rho}^{[j]}_{t}\rangle\rangle=\int_{-\infty}^{\infty} \dd D \left[\mathcal{L}_D +\lambda\mathcal{D}_A\right]|\rho^{[j]}_{t}(D)\rangle\rangle .
\end{equation}
These conditions lead to a set of master equations describing the dynamics of the system after the detector variable $D$ is integrated out. In the next section we explicitly obtain these master equations in the 0th and 1st order of the perturbation scheme.
\par

Although we have assumed that $\mathcal{L}_D$ is time-independent in arriving at these results, our analysis remains valid if $\mathcal{L}_D$ depends on the slow time variable $\tau_1=t$.

\subsubsection{Master equations}
Setting $j=0$ in Eq.~\eqref{secular_condn_sep_time} and using the 0th order solution Eq.~\eqref{0th_sep_time}, we obtain
\begin{equation}\label{0th_order_master_eq}
     \partial_t |\bar{\rho}^{[0]}_{t}\rangle\rangle=\left[\bar{\mathcal{L}}_0 + \lambda\mathcal{D}_A \right]|\bar{\rho}^{[0]}_{t}\rangle\rangle,
\end{equation}
where
\begin{equation}\label{eq:Lbar0}
    \bar{\mathcal{L}}_0=\int \dd D \mathcal{L}_D G(D).
\end{equation}
In Eq.~\eqref{0th_order_master_eq} we have
used the fact that $\mathcal{D}_A$ is a diagonal matrix and is independent of the detector variable $D$.
The matrix $\bar{\mathcal{L}}_0$ is the Fock-Liouville representation of the superoperator $\mathcal{L}_0$ of the main text.

Next, setting $k=1$ in Eq.~\eqref{kth_sep_time} and using Eqs.~\eqref{0th_sep_time}, (\ref{0th_order_master_eq}) and (\ref{eq:Lbar0}), we get
\begin{equation}\label{1st_sep_time}
    |\rho_t^{[1]}(D)\rangle\rangle= G(D)|\bar{\rho}_t^{[1]}\rangle\rangle-\Gamma^{-1} \mathcal{F}^{+}\mathcal{L}_DG(D)|\bar{\rho}_t^{[0]}\rangle\rangle.
\end{equation}
Now substituting this expression into Eq.~\eqref{secular_condn_sep_time} with $j=1$, we arrive at
\begin{equation}
\label{1st_order_master_eq}
    \partial_t |\bar{\rho}^{[1]}_{t}\rangle\rangle=\left[\bar{\mathcal{L}}_0 + \lambda\mathcal{D}_A \right]|\bar{\rho}^{[1]}_{t}\rangle\rangle +\Gamma^{-1}\bar{\mathcal{L}}_1|\bar{\rho}^{[0]}_{t}\rangle\rangle,
\end{equation}
where
\begin{equation}
     \bar{\mathcal{L}}_1=-\int \dd D \ \mathcal{L}_D \mathcal{F}^{+} \mathcal{L}_D G(D).
\end{equation}

We can take a step further and rewrite this first order correction in terms of time integrals of correlation functions
\begin{equation}
\label{eq:1st_order_correlation}
    (\bar{\mathcal{L}_1})_{bb',aa'} = \sum_{cc'} \int_0^\infty \dd t~\left[ \langle (\mathcal{L}_D)_{bb',cc'}(t) (\mathcal{L}_D)_{cc',aa'}(0) \rangle_{\pi_{aa'}}^{cc'} - \langle (\mathcal{L}_D)_{bb',cc'} \rangle_{\pi_{cc'}} \langle (\mathcal{L}_D)_{cc',aa'} \rangle_{\pi_{aa'}} \right].
\end{equation}
Here $\langle \cdot \rangle_\phi$ denotes an ensemble average over the probability distribution $\phi(D)$, and $\langle \psi(t) \xi(0) \rangle_\phi^{cc'}$ denotes the ensemble average of the function $\psi(D_t) \xi(D_0)$, with $D_0$ initially sampled from $\phi(D_0)$, and the superscript $cc'$ indicates that $D_t$ is obtained by evolving $D_0$ for a time $t$ under the dynamics generated by $\mathcal{F}_{cc'}$. The integral in Eq. \eqref{eq:1st_order_correlation} converges because the integrand decays exponentially fast to zero, by properties of OU dynamics.

Combining Eqs.~\eqref{0th_order_master_eq} and Eq.~\eqref{1st_order_master_eq} we get a master equation for the system dynamics alone, to $\mathcal{O}(\epsilon^1)$
\begin{equation}\label{eq:0th+1st_order_master_eq}
    \partial_t|\bar{\rho}_t\rangle\rangle=\left[\bar{\mathcal{L}}_0 +\lambda\mathcal{D}_A+\gamma^{-1} \bar{\mathcal{L}}_1\right]|\bar{\rho}_t\rangle\rangle ,
\end{equation}
where we have used $|\bar{\rho}_t\rangle\rangle =|\bar{\rho}^{[0]}_t\rangle\rangle + \epsilon |\bar{\rho}^{[1]}_t\rangle\rangle$ and $\epsilon \Gamma^{-1} = \gamma^{-1}$. This result corresponds to Eq.~(\ref{eq:Markovian-ME}) of the main text.

The joint system-detector state, to $\mathcal{O}(\epsilon^1)$, is obtained from Eqs.~(\ref{0th_sep_time}) and (\ref{1st_sep_time})
\begin{equation}
\label{eq:joint_rho_detector}
    |\rho_t(D)\rangle\rangle = \left[ I - \gamma^{-1}\mathcal{F}^+ \mathcal{L}_D \right] G(D) |\bar\rho_t \rangle\rangle,
\end{equation}
where $I$ is the identity matrix and $|\bar\rho_t \rangle\rangle$ evolves under Eq.~(\ref{eq:0th+1st_order_master_eq}).

\section{General numerical method for solving the QFPME}
\label{sec:general-numerics}
Before providing detailed analytical calculations for the toy models based on the separation of time-scale approaches discussed above, we briefly outline how Eq.~(\ref{eq:MAIN-RESULT}) was solved numerically for these models. While we considered two-level systems, generalizing the method to higher dimensions is straightforward.

To numerically find the steady state of Eq.~(\ref{eq:MAIN-RESULT}) for the toy models, we expand the density operator in terms of the generalized Hermite polynomials [defined in Eq.~(\ref{eq:hermitesigm})] as
\begin{equation}
\hat{\rho}_t(D) = \sum_{n=0}^{N-1} M_n \frac{He_n^{[\sigma]}(D)}{\sqrt{\sigma^n n!}} \frac{e^{-D^2/2\sigma}}{\sqrt{2\pi\sigma}}, \hspace{2cm} M_n =
\begin{pmatrix}
a_n & c_n \\
c_n^* & b_n
\end{pmatrix},
\end{equation}
\noindent where the sum is truncated to $N$ terms, the matrix $M_n$ is written in the $\{\ket{0},\ket{1}\}$ basis, and $a_n$, $b_n$, and $c_n$ are expansion coefficients for the respective elements of the density matrix. %, with $c_n^*$ being the complex conjugate of $c_n$. We use $\sigma = \gamma/8\lambda$ such that the diffusion constant of the Ornstein-Uhlenbeck process [see Eq.~(1)] is incorporated into the generalized Hermite polynomials. 
For the classical toy model, $c_n$ can be put to zero. Inserting this expansion in Eq.~(\ref{eq:MAIN-RESULT}), multiplying with $He_m^{[\sigma]}/\sqrt{\sigma^m m!}$, and integrating over $\int_{-\infty}^\infty dD$ results in a relation
\begin{equation}
0 = \boldsymbol{f}_m(\{a_n,b_n,c_n\}_{n=0}^{N-1}),
\label{eq:recurrence}
\end{equation}
\noindent where $\boldsymbol{f}_m(\cdot)$ is a vector valued function of the expansion coefficients.  Together with the normalization condition $1 = \int_{-\infty}^\infty dD \trace\{\hat{\rho}_t(D)\} = a_0 + b_0$, we numerically find the expansion coefficients by rewriting Eq.~(\ref{eq:recurrence}) as a matrix equation.

To explicitly find the function $\boldsymbol{f}_m$ for threshold feedback, the following identity was used
\begin{equation}
\int_{0}^\infty \frac{He_m^{[\sigma]}(x)}{\sqrt{\sigma^m m!}} \frac{He_n^{[\sigma]}(x)}{\sqrt{\sigma^n n!}} \frac{e^{-x^2/2\sigma}}{\sqrt{2\pi\sigma}} dx =
\begin{cases}
1/2, \hspace{0.5cm} n=m, \\
0, \hspace{0.85cm} n+m \hspace{0.1cm} \text{even},\\
C_{nm}, \hspace{0.35cm} n+m \hspace{0.1cm} \text{odd},
\end{cases}
\hspace{1cm}
C_{nm} = \frac{(-1)^{(n+m-1)/2}m!!(n-1)!!}{\sqrt{2 \pi n! m!}(m-n)},
\end{equation}
\noindent where $C_{nm}$ is given for even $n$ and odd $m$, and we note that $(-1)!!=1$.

\section{Classical toy model}
\label{sec:classical-toy}
\subsection{Recovering a classical model}
\label{sec:classical-model-rateEq}
Starting from Eq.~(\ref{eq:MAIN-RESULT}) of the main text, we can recover an effective classical model by considering a Liouville superoperator ${\mathcal L}(D)$ that does not connect diagonal elements (populations) with off-diagonal elements (coherences), in the basis of the measured operator $\hat A$. In this case, the coherences decouple from the populations and decay exponentially with time, provided the spectrum of $\hat{A}$ is non-degenerate.  We will illustrate this approach below.  An alternative approach (which we do not pursue here) is to consider a model where the coherences are destroyed by a large measurement backaction or an additional decoherence term.% In this case the steady state coherences scale as $1/\lambda$ and therefore vanish as $\lambda\rightarrow\infty$.

To obtain a Liouville superoperator with the above-mentioned property, we assume that the system Hamiltonian $\hat H$ commutes with the measured operator $\hat A$, allowing us to write $\hat H|a\rangle = \omega_a |a\rangle$ and $\hat A|a\rangle = \xi_a |a\rangle$, and we consider a superoperator $\mathcal{L}(D)$ of the form
\begin{equation}
    \label{eq:Liouville_D}
    \mathcal{L}(D) \hat{\rho}_t(D) = -i\left[ H(D), \hat{\rho}_t(D) \right] + \sum_{aa'} M_{aa'}(D) \mathcal{D}\left[ |a \rangle \langle a' | \right] \hat{\rho}_t(D),
\end{equation}
where the $M_{aa'}$'s are the coefficients of jump operators in the basis $\{ |a\rangle \}$. Eq.~(\ref{eq:MAIN-RESULT}) then determines the dynamics for the populations and coherences in this basis. The populations evolve under an effective rate equation
\begin{equation}
    \label{eq:pop_rate_matrix}
    \partial_t \rho_{aa}(D) = \sum_{a'\ne a} M_{aa'}(D) \rho_{a'a'}(D) - \sum_{a'\ne a} M_{a'a}(D) \rho_{aa}(D) +\gamma \mathcal{F}_{aa} \rho_{aa}(D) \equiv \sum_{a'} W_{aa'}(D) \rho_{a'a'}(D) +\gamma \mathcal{F}_{aa} \rho_{aa}(D),
\end{equation}
where off-diagonal terms of the rate matrix are $W_{aa'} = M_{aa'}$, diagonal terms are $W_{aa} = -\sum_{a'\ne a} M_{a'a}$ and $\mathcal{F}_{aa'} $ is defined by Eq.~(\ref{eq:ou}). The coherences obey ($a\neq a'$)
\begin{equation}
    \label{eq:coh_decay}
    \partial_t \rho_{aa'}(D) = -i\left( \omega_a(D) - \omega_{a'}(D) \right) \rho_{aa'}(D) -\frac{\lambda}{2}(\xi_a - \xi_{a'})^2 \rho_{aa'}(D) + \frac{W_{aa}(D) + W_{a'a'}(D)}{2} \rho_{aa'}(D) +\gamma \mathcal{F}_{aa'} \rho_{aa'}(D).
\end{equation}
Note that the decay rate is proportional to the flux of probability out of state $a$ and $a'$ and a contribution due to the measurement. In the long time limit the coherences decay to zero.

We see from Eq.~(\ref{eq:pop_rate_matrix}) that the backaction term effectively drops out of the equations of motion for the populations. As a result, to use separation of time scales, we no longer need to assume that $\gamma\gg\lambda$ (as required in the general quantum setting, see main text) thus we can use arbitrary values of $\lambda / \gamma$. In a classical setting, measurements do not inherently disturb a system, hence $\lambda$ only determines the information per unit time that is gathered about the physical system.

In the remainder of this subsection we focus on the evolution of the the detector distribution $\rho_{aa}(D)$ that corresponds to the populations, and we ignore the distribution $\rho_{a a'}(D)$ corresponding to coherences.  This effectively means that we will work in an $N$-dimensional space rather than an $N^2$-dimensional space.  Equivalently, we will now work with classical probability vectors rather than quantum density matrices. Following the notational convention of the main text, see comments after Eq.~(\ref{eq:MAIN-RESULT}), we will use $\vec{\rho}(D,t)$ to denote the joint classical probability distribution of the system and detector, and $\vec{\rho}(t)$ to denote the probability distribution of the system alone.
Similar comments apply to the steady state distributions $\vec{\rho}_{ss}(D)$ and $\vec{\rho}_{ss}$.

At leading order of approximation, the equations of motion for the populations $\rho_{aa}$ follow from Eqs.~(\ref{0th_order_master_eq}) and (\ref{eq:Lbar0}), resulting in the classical rate equation
\begin{equation}
    \label{eq:0th_rate_eq}
    \partial_t \vec{\rho}\,^{[0]} = \tilde{W}_0 \vec{\rho}\,^{[0]},
\end{equation}
where
\begin{equation}
    \vec{\rho}\,^{[0]} = \left( \rho_{11}^{[0]}, \rho_{22}^{[0]} , \cdots \rho_{NN}^{[0]} \right)^T,
\end{equation}
is a vector that contains all the populations, and $\tilde{W}_0$ is an $N\times N$ matrix with elements
\begin{equation}\label{eq:tW0def}
    (\tilde{W}_0)_{aa'} = \int \dd D \,  W_{aa'}(D) \pi_{a'a'}(D) .
\end{equation}
By Eq.~(\ref{eq:DA}), the elements of $\mathcal{D}_A$ that operate on  populations $\rho_{aa}$ vanish, hence the second term $\lambda\mathcal{D}_A$ on the right side of Eq.~(\ref{0th_order_master_eq}) does not contribute to Eq.~(\ref{eq:0th_rate_eq}).

At the next order of approximation, Eq.~(\ref{1st_order_master_eq}) gives us
\begin{equation}\label{eq:1st_rate_eq}
    \partial_t \vec{\rho}\,^{[1]} = \tilde{W}_0 \vec{\rho}\,^{[1]} +\Gamma^{-1} \tilde{W}_1 \vec{\rho}\,^{[0]} ,
\end{equation}
where $\vec{\rho}\,^{[1]}$ is defined similarly to $\vec{\rho}\,^{[0]}$, and the $N\times N$ matrix $\tilde{W}_1$ is given by
\begin{equation}
    \label{eq:classical_1st_order}
    \tilde{W}_1 = - \int \dd D \, W(D) \, \mathcal{F}_c^+ \, W(D) \, G_c(D).
\end{equation}
Here $G_c$ is the $N\times N$ diagonal matrix
\begin{equation}
    G_c(D)=\sum_{i} |aa\rangle\rangle\pi_{aa}(D)\langle\langle aa|,
\end{equation}
whose elements are obtained by keeping only those elements of $G$, see Eq.~(\ref{eq:Gdef}), that correspond to populations and not to coherences.
Similarly, $\mathcal{F}_c^+$ is the $N\times N$ diagonal matrix obtained by keeping only those elements of $\mathcal{F}^+$ that correspond to populations and not to coherences. The classical analog of Eq.~(\ref{eq:1st_order_correlation}) can be written as:
\begin{equation}
    \label{eq:classical_1st_order_elements}
    (\tilde{W}_1)_{aa'} =  \sum_b \int_0^\infty \dd t  \left[\langle W_{ab}(t)W_{ba'}(0) \rangle_{\pi_{a'a'}}^{bb} - \langle W_{ab} \rangle_{\pi_{bb}} \langle W_{ba'} \rangle_{\pi_{a'a'}} \right].
\end{equation}

Combining results and setting $\vec\rho = \vec{\rho}\,^{[0]} + \epsilon \vec{\rho}\,^{[1]}$, we obtain a master equation valid to first order in $\epsilon$:
\begin{equation}\label{eq:mevecrho}
        \partial_t \vec{\rho} = \tilde{W}_0 \vec{\rho} +\gamma^{-1}\tilde{W}_1 \vec{\rho} ,
\end{equation}
where we have used $\epsilon\Gamma^{-1}=\gamma^{-1}$.
This result corresponds to Eq.~(\ref{eq:0th+1st_order_master_eq}), applied to our classical model. The joint system-detector distribution for the classical case, to $\mathcal{O}(\epsilon^1)$, follows from Eqs. (\ref{0th_sep_time}) and (\ref{1st_sep_time}):
\begin{equation}
   \vec{\rho}(D,t)=\left[ I_c -\gamma^{-1}\mathcal{F}_c^{+}W(D) \right] G_c(D)\vec{\rho}(t) ,
\end{equation}
where $I_c$ is the $N\times N$ identity operator and $\vec{\rho}(t)$ obeys Eq.~\eqref{eq:mevecrho}. This is the classical analogue of Eq.~(\ref{eq:joint_rho_detector}).

\subsection{Analytical Calculations}
\subsubsection{Perturbative solution to QFPME for classical toy model}
As discussed in Sec.~\ref{sec:classical-model-rateEq}, for the classical toy model we can write down rate equations in the classical subspace, that is the subspace of populations only, without coherences.
In what follows, we first use the results of Sec.~\ref{sec:classical-model-rateEq} to solve for $\vec{\rho}_{ss}$, and from that result we determine $\vec{\rho}_{ss}(D)$.

For the particular model of threshold feedback we have
\begin{equation}
\label{eq:classical_feedback}
W(D)=W_{+}\theta(D) + W_{-}(1-\theta(D)) ,
\end{equation}
where
\begin{equation}
W_{+}=\Gamma\begin{pmatrix}
-(n_B+1) & n_B \\
(n_B +1) & -n_B
\end{pmatrix} 
,\quad \quad
W_{-}=
\Gamma \begin{pmatrix}
-n_B & (n_B +1) \\
n_B & -(n_B+1)
\end{pmatrix}.
\end{equation}
Here, Eq.~\eqref{eq:classical_feedback} is equivalent to the Eq.~(\ref{eq:Classical-Liouville-D}) of the main text and $W_{+}$ and $W_{-}$ are rate matrices that correspond to the classical subspace of the superoperators $\mathcal{L}_{+}$ and $\mathcal{L}_{-}$ respectively. Matrices $G_c(D)$ and $\mathcal{F}^{+}_c$ are given by
\begin{equation}
G_c(D)=\begin{pmatrix}
\pi_{00}(D) & 0 \\
0 & \pi_{11}(D)
\end{pmatrix} ,\quad\quad
\mathcal{F}^{+}_c=\begin{pmatrix}
\mathcal{F}_{00}^{+} & 0 \\
0 & \mathcal{F}_{11}^{+}
\end{pmatrix}.
\end{equation}
Where $\pi_{00}$ and $\pi_{11}$ are Gaussian distributions centered at $-1$ and $1$, respectively. They follow from Eq.~(\ref{eq:stationary-OU-distr}) with the particular choice for the measurement operator $\hat{A} = \hat{\sigma}_z$ with $\xi_0 = -1$ and $\xi_1 = 1$ as introduced in the main text.  From these expressions we can evaluate the matrices $\tilde{W}_0$ and  $\tilde{W}_1$ defined in Sec.~\ref{sec:classical-model-rateEq}.
Eq.~(\ref{eq:tW0def}) gives
\begin{equation}\label{eq:tW0}
\tilde{W}_0 = \Gamma(n_B + \eta)
\begin{pmatrix}
-1 & 1 \\
1 & -1
\end{pmatrix},
%\quad,\quad
%\tilde{W}_1 = \Gamma^2(m-z-kn_B)\begin{pmatrix}
%-1 & 1 \\
%1 & -1
%\end{pmatrix}
\end{equation}
where 
\begin{equation}
\label{eq:eta}
\eta = \frac{1}{2}[1-\erf (2\sqrt{\lambda/\gamma})],
\end{equation}
is the error probability.

Equation (\ref{eq:tW0}) leads to the following normalized steady state solution of Eq.~(\ref{eq:0th_rate_eq})
\begin{equation}\label{eq:rhovecss0}
\vec{\rho}\,^{[0]}_{ss}=
\begin{pmatrix}
1/2 \\ 1/2
\end{pmatrix}.
\end{equation}
Equation (\ref{eq:classical_1st_order}) gives
\begin{equation}\label{eq:tW1}
\tilde{W}_1= \Gamma^2(m-z-kn_B)\begin{pmatrix}
-1 & 1 \\
1 & -1
\end{pmatrix},
\end{equation}
with
\begin{eqnarray}
\label{eq:const_m}
m &=& \int \dd D \, \theta(D)\mathcal{F}^{+}_{00}\theta(D)\pi_{00}(D) = \int \dd D \, \phi(D) \mathcal{F}_{11}^{+}\phi(D) \pi_{11}(D),
\\
\label{eq:const_z}
z &=& \int \dd D \, \theta(D)\mathcal{F}^{+}_{11}\theta(D)\pi_{00}(D) = \int \dd D \, \phi(D)\mathcal{F}^{+}_{00}\phi(D)\pi_{11}(D),
\\
\label{eq:const_k}
k &=& \int \dd D \, \theta(D)\mathcal{F}^{+}_{11}\pi_{00}(D)= \int \dd D \, \phi(D)\mathcal{F}_{00}^{+} \pi_{11}(D),
\end{eqnarray}
where $\phi(D)=1-\theta(D)$.\par 
Setting the left side in Eq.~\eqref{eq:mevecrho} to zero and using Eqs.~(\ref{eq:tW0}), (\ref{eq:tW1}), we obtain the steady state solution
\begin{equation}\label{eq:rhovecss}
\vec{\rho}_{ss} =
\begin{pmatrix}
1/2 \\ 1/2
\end{pmatrix},
\end{equation}
which together with Eq.~(\ref{eq:rhovecss0}) implies
\begin{equation}\label{eq:rhovecss1}
\vec{\rho}\,^{[1]}_{ss} =
\begin{pmatrix}
0 \\ 0
\end{pmatrix}.
\end{equation}

We now use these results to construct the joint steady state probability distribution $\vec{\rho}_{ss}(D)$.
Combining Eq.~(\ref{eq:rhovecss0}) and the classical analogue of Eq.~(\ref{0th_sep_time}) give
the 0th order joint steady state distribution
\begin{equation}
\vec{\rho}\,^{[0]}_{ss}(D)=\frac{1}{2} 
\begin{pmatrix}
\pi_{00}(D) \\ \pi_{11}(D)
\end{pmatrix} .
\end{equation}
Similarly, Eqs.~(\ref{1st_sep_time},\ref{eq:rhovecss0},\ref{eq:rhovecss1}) give
\begin{equation}
\vec{\rho}\,^{[1]}_{ss}(D)=-\Gamma^{-1} \mathcal{F}_c^{+}W(D)G_c(D)\begin{pmatrix}
1/2 \\
1/2
\end{pmatrix}.
\end{equation}
If we now define for convenience a function
\begin{equation}
h(D)=\frac{1}{2}\Bigl[W_{10}(D)\pi_{00}(D)-W_{01}(D)\pi_{11}(D)\Bigr],
\end{equation}
then the joint system detector steady state distribution can be written, to first order, as
\begin{equation}
\label{eq:pss_classical}
\vec{\rho}_{ss}(D)=\frac{1}{2}\begin{pmatrix}
\pi_{00}(D) \\
\pi_{11}(D)
\end{pmatrix}
+\epsilon\Gamma^{-1} \begin{pmatrix}
\mathcal{F}^{+}_{00} h(D)\\
- \mathcal{F}^{+}_{11} h(D)
\end{pmatrix}.
\end{equation}
Upon integrating this expression over the variable $D$, we recover Eq.~(\ref{eq:rhovecss}).

\subsubsection{Power calculation}

\begin{figure}[H]
	\centering
	\includegraphics[scale=0.4]{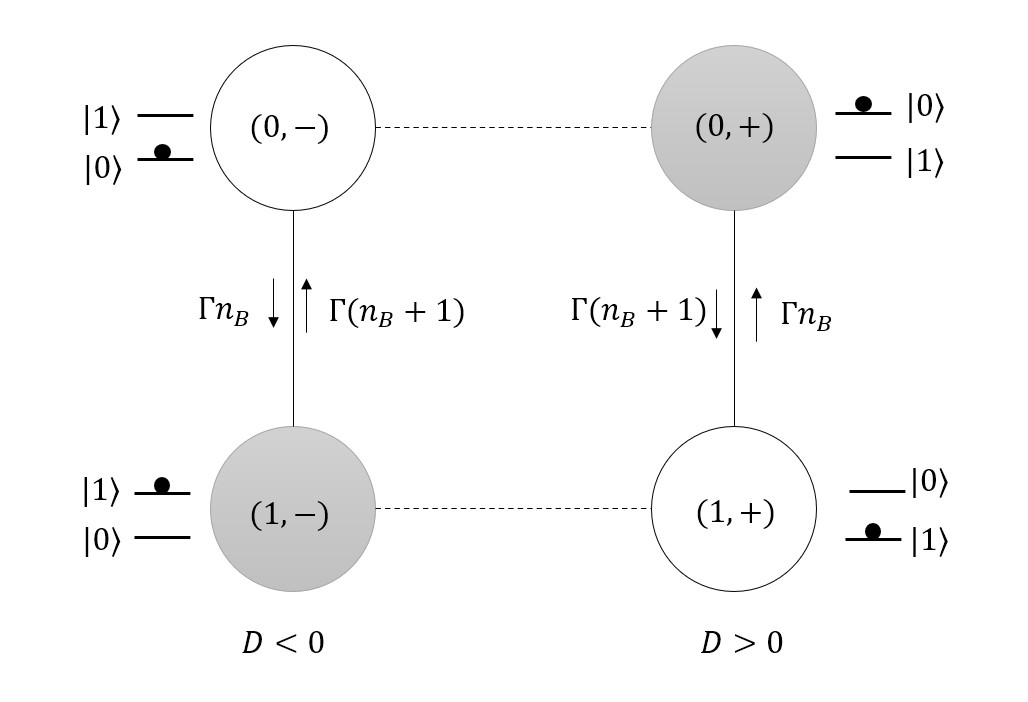}
	\caption{Coarse grained classical toy model. States are shown in the circles. White colored states correspond to the states where detector variable $D$ captures the system state properly. Grey colored states corresponds to the states where the detector variable $D$ fails to capture the actual state of the system. Edges shown with solid lines correspond to bath assisted transitions and the corresponding rates are shown on the sides of the edges. Edges marked with dotted lines correspond to changes in sign of the detector variable $D$ due to its diffusive dynamics. Note that the feedback is instantly applied when $D$ changes sign. Each of the $(1,-)\rightarrow(1,+)$ and $(0,+)\rightarrow(0,-)$ transitions correspond to extraction of $\Delta$ work from the system. Hence in each counter-clockwise cycle $2\Delta$ work is extracted.}
	\label{fig:Classical_toy_model}
\end{figure}

Now we move to the calculation of the power extracted or dissipated by the feedback. There are two main sources of dissipation in the system: weak measurements leading to limited information and incorrect application of feedback, and slow detector dynamics that lag with respect to the system dynamics. The first order correction for the power using separation of timescales in $\gamma$ captures the latter source of dissipation.

To capture both effects we introduce a discrete state model containing four states, $\{(0,-),(1,-),(1,+),(0,+)\}$, where the two entries of each state describe the system and the sign of the detector variable $D$; see Fig. \ref{fig:Classical_toy_model}. The states $(0,-)$ and $(1,+)$ are ones in which the detector state correctly represents the system state, whereas $(1,-)$ and $(0,+)$ describe erroneous states of the detector. 
These four states form a cyclic network in which the transitions $(0,-)\leftrightarrow(1,-)$ are governed by the rate matrix $W_{-}$, the transitions $(0,+)\leftrightarrow(1,+)$ are governed by the rate matrix $W_{+}$, and the transitions $(0,+)\leftrightarrow(0,-)$ and $(1,-)\leftrightarrow(1,+)$ arise from the diffusive dynamics of the detector variable $D$. 
In the steady state, the probabilities of this coarse grained model can be written as
\begin{equation}
\begin{pmatrix}
P^{ss}_{0,+} \\
P^{ss}_{1,+}
\end{pmatrix}=
\int_{-\infty}^{\infty} \dd D \ \theta(D) \vec{\rho}_{ss}(D),\quad\quad \begin{pmatrix}
P^{ss}_{0,-} \\
P^{ss}_{1,-}
\end{pmatrix}=
\int_{-\infty}^{\infty} \dd D \ (1- \theta(D)) \vec{\rho}_{ss}(D),
\end{equation}
where $\vec{\rho}_{ss}(D)$ is given by Eq.~\eqref{eq:pss_classical}.

The desired mode of operation corresponds to counterclockwise transitions around the network shown in Fig.~\ref{fig:Classical_toy_model}, and in every cycle $2\Delta$ work is extracted, where $\Delta$ is the energy separation of the states $|0\rangle$ and $|1\rangle$. In the steady state, the probability currents are equal across all four edges and can be calculated from any of the edges. Here we define power as $P=2\Delta J_{ss}=2\Delta J_{(1,+)\rightarrow(0,+)}=2\Delta\left[ (W_+)_{01}P_{(1,+)}-(W_+)_{10}P_{(0,+)}\right]$. We then obtain
\begin{equation}
P =\Delta\Gamma\left[n_B(1-\eta) -(n_B +1)\eta\right]- \epsilon 2 \Delta \left[n_B   \int \dd D\theta(D)\mathcal{F}^+_{11}h(D) + (n_B + 1) \int \dd D \theta(D)\mathcal{F}_{00}^+h(D)\right],
\end{equation}
which can be written as
\begin{equation}
\label{eq:power_classical_MD}
P =\Delta\Gamma\left[n_B(1-\eta) -(n_B +1)\eta\right]- \gamma^{-1} \Delta\Gamma^2\left[(2 n_B +1)(m+z+kn_B)\right],
\end{equation}
with $m, z$ and $k$ defined by Eqs.~(\ref{eq:const_m},\ref{eq:const_z},\ref{eq:const_k}) [for an alternative expression, see Eq.~\eqref{eq:map}]. We see that even in the 0th order in the detector dynamics, the extracted power decreases with $\eta$, reflecting the negative effects of weak measurements that gather insufficient information from the system.

\subsubsection{Strong measurement approximation}
When the detector is accurate $(\lambda\gg\gamma) $ the distributions $\pi_{00}$ and $\pi_{11}$ become very localized.
In that situation $\eta \approx0$, and the function $h(D)$ can be approximated as $h(D)\approx \Gamma(n_B +1)\left[\pi_{00}(D) -\pi_{11}(D)\right] $. We also have 
\begin{equation}
\mathcal{F}_{00}^{+}\pi_{11}(D)=-\int_0^{\infty}\dd s [\pi(D;2e^{-s}-1)-\pi_{00}(D)]    , \ \ \mathcal{F}_{11}^{+}\pi_{00}(D)=-\int_0^{\infty}\dd s [\pi(D;1-2e^{-s})-\pi_{11}(D)],
\end{equation}
where $\pi(D;y)$ denotes the stationary distribution of an OU process centered at $y$. Eq. (\ref{eq:const_k}) can now be approximated as
\begin{equation}
k=\int_{-\infty}^{\infty} \dd D \theta(D)\mathcal{F}^{+}_{11}\pi_{00}(D) = -\int_{0}^{\infty} \dd s \int_{0}^{\infty} \dd D [\pi(D;1-2e^{-s})-\pi_{11}(D)]\approx -\int_{0}^{\ln{2}}\dd s (-1)=\ln{2} ,
\end{equation}
and similarly in this approximation $m\approx0$ and $z\approx0$. Using these we can write down the expression for power at very strong measurement limit as
\begin{equation}
P=n_B \Gamma\Delta\ \left[ 1 - \gamma^{-1}\Gamma(2n_B +1)\ln{2}\right],
\end{equation}
where the negative contribution in the first order reflects the fact that we get less extracted power when the detector lag cannot be neglected.

\subsubsection{Fluctuation theorem}
\label{sec:fluctuation-thm}

Here we derive Eq.~(\ref{eq:fluctuation-thm}) in the main text. We may write Eq.~\eqref{eq:tW0} as
\begin{equation}
\tilde{W}_0 =
\begin{pmatrix}
-\tilde{W}_{10}^{(-)} & \tilde{W}_{01}^{(-)} \\
\tilde{W}_{10}^{(-)} & -\tilde{W}_{01}^{(-)}
\end{pmatrix}+\begin{pmatrix}
-\tilde{W}_{10}^{(+)} & \tilde{W}_{01}^{(+)} \\
\tilde{W}_{10}^{(+)} & -\tilde{W}_{01}^{(+)}
\end{pmatrix},
\end{equation}
where $\tilde{W}_{aa'}^{(\nu)}$ is the transition rate from state $a'$ to $a$ with level configuration $\nu=\pm$, for which $-$ corresponds to the configuration with $\ket{0}$ as ground state and $+$ to the one with $\ket{1}$ as ground state. The rates are given by
\begin{equation}
\tilde{W}_{01}^{(-)} = \eta \Gamma [n_B(\Delta)+1],\hspace{.5cm}
\tilde{W}_{10}^{(-)} = (1-\eta) \Gamma n_B(\Delta),
\hspace{.5cm}
\tilde{W}_{01}^{(+)} = (1-\eta) \Gamma n_B(\Delta),
\hspace{.5cm}
\tilde{W}_{10}^{(+)} = \eta \Gamma [n_B(\Delta)+1],
\end{equation}
\noindent where $\eta$ is the error probability defined in Eq.~\eqref{eq:eta}. The following local detailed balance relation holds,
\begin{equation}
\ln\left( \frac{\tilde{W}_{01}^{(\pm)}}{\tilde{W}_{10}^{(\pm)}} \right) = \mp \left[ \frac{\Delta}{k_B T} - \ln\left( \frac{1-\eta}{\eta} \right) \right],
\end{equation}
\noindent where $k_B$ is the Bolzmann constant. This is in line with the postulated relation for local detailed balance for Maxwell demon feedback given in Ref.~\cite{Esposito-EPL-2012}, and we now follow this reference to derive the fluctuation theorem.

We begin by defining a forward trajectory $X = \{ (t_j, \nu_j, S_{j-1} \rightarrow S_j) \}_{j=1}^n$, specifying that at time $t_j$, with state configuration $\nu_j$, the system transit from state $S_ {j-1}$ to $S_j$. A time reversed trajectory $X^{\rm tr}$ can be defined analogously. In steady state, we find the detailed fluctuation theorem
\begin{equation}
\frac{P(X^{\rm tr})}{P(X)} = e^{M(X) \left[ \frac{\Delta}{k_B T} - \ln\left( \frac{1-\eta}{\eta} \right) \right]},
\end{equation}
\noindent where $P(X)$ denotes the probability for observing trajectory $X$ and $M(X)$ the number of extracted energy quanta for trajectory $X$. Defining the probability of extracting $m$ energy quanta as $P(m) = \sum_{k: M(X_k)=m} P(X_k) $, summing over all trajectories with $m$ extractions, we find Eq.~(\ref{eq:fluctuation-thm}) in the main text.

\subsection{Alternative power calculation}
\label{sec:classical-toy-power}

In this section, we provide an alternative calculation for the power in the classical toy model that can be used to access also the power fluctuations. To this end, we employ the method of full counting statistics, following Ref.~\cite{schaller-book}. We start with the density matrix formalism to illustrate the general parts of this derivation. The specific calculations are then carried out in the vector notation used above. The probability of having $n$ transferred energy quanta from reservoir to two-level system after time $t$ is given by $P_t(n) = \int_{-\infty}^\infty dD \trace\{\hat{\rho}_t(D,n)\}$, where $\hat{\rho}_t(D,n)$ is the number-resolved system-detector density operator. The joint state of system and detector $\hat{\rho}_t(D) = \sum_n \hat{\rho}_t(D,n)$. The discrete Fourier transform of the number-resolved density operator reads $\hat{\rho}_t(D,\chi) = \sum_n \hat{\rho}_t(D,n) e^{in\chi}$, where we introduced a counting field $\chi$. We can now write the counting field-resolved version of Eq.~(\ref{eq:MAIN-RESULT}) as
\begin{equation}
\partial_t \hat{\rho}_t(D,\chi) = \mathcal{L}(D,\chi) \hat{\rho}_t(D,\chi) + \lambda \mathcal{D}[\hat{A}] \hat{\rho}_t(D,\chi) - \gamma \partial_D \mathcal{A}(D) \hat{\rho}_t(D,\chi) + \frac{\gamma^2}{8 \lambda} \partial_D^2 \hat{\rho}_t(D,\chi),
\end{equation}
\noindent with Liouvillian $\mathcal{L}(D,\chi) = [1-\theta(D)]\mathcal{L}_{-}(\chi)+\theta(D)\mathcal{L}_{+}(\chi)$, where $\mathcal{L}_{-}(\chi) = \Gamma n_B(\Delta)\mathcal{D}_{+}^\chi[\hat{\sigma}^\dagger] + \Gamma[n_B(\Delta)+1]\mathcal{D}_{-}^\chi[\hat{\sigma}]$ and $\mathcal{L}_{+}(\chi)=\Gamma[n_B(\Delta)+1]\mathcal{D}_{-}^\chi[\hat{\sigma}^\dagger] + \Gamma n_B(\Delta) \mathcal{D}_{+}^\chi[\hat{\sigma}]$, with counting field dependent dissipators $\mathcal{D}_{\pm}^\chi[\hat{O}]\hat{\rho} = e^{\pm i\chi}\hat{O}\hat{\rho}\hat{O}^\dagger - \frac{1}{2}\{\hat{O}^\dagger\hat{O},\hat{\rho}\}$, and $\hat{\sigma}=|0\rangle\langle 1|$. We further define the moment generating function
\begin{equation}
\psi_t(\chi) \equiv \int_{-\infty}^\infty dD \trace \left\{ \hat{\rho}_t(D,\chi) \right\},
\end{equation}
\noindent such that the average number of transferred quanta reads $\langle n \rangle = -i \partial_\chi \psi_t(\chi)|_{\chi=0}$. As $t \rightarrow \infty$, the steady state current of quanta is given by
\begin{equation}
\frac{\langle n \rangle}{t} = -i \int_{-\infty}^\infty dD \trace \left\{ \left[ \partial_\chi \mathcal{L}(D,\chi) \right]_{\chi=0} \hat{\rho}_{\rm ss}(D) \right\},
\label{eq:classical-power-FCS}
\end{equation}
\noindent where $\hat{\rho}_{\rm ss}(D)$ is the steady state solution to Eq.~(\ref{eq:MAIN-RESULT}). The steady state power now becomes $P = \Delta \langle n \rangle / t$. Equation \eqref{eq:classical-power-FCS} may be computed numerically, using the method outlined in Sec.~\ref{sec:general-numerics}, or, assuming a separation of time-scales, analytically as outlined below.

Following the method outlined in Sec.~\ref{sec:threshfeed}, we find (using the vector notation introduced in Sec.~\ref{sec:classical-model-rateEq})
\begin{equation}
\label{eq:l0toy}
\tilde{W}_0(\chi) = -\Gamma(n_B+\eta)
I+\Gamma\left[e^{i\chi\Delta}n_B(1-\eta)+e^{-i\chi\Delta}(n_B+1)\eta\right]\sigma_x,
\end{equation}
where $\sigma_x$ denotes the Pauli $x$-matrix, $I$ the identity in two dimensions, and we abbreviated $n_B\equiv n_B(\Delta)$. For the first order correction, we obtain
\begin{equation}
\label{eq:lcorrtoy}
\begin{aligned}
\tilde{W}_1(\chi)/\Gamma^2 &= IC_1-I\left[n_Be^{i\chi\Delta}-(n_B+1)e^{-i\chi\Delta}\right]\left\{n_Be^{i\chi\Delta}[C_2+C_0(1-\eta)]-(n_B+1)e^{-i\chi\Delta}\left[C_2-C_0\eta\right]\right\}\\&+\sigma_xC_1\left[n_Be^{i\chi\Delta}-(n_B+1)e^{-i\chi\Delta}\right]-\sigma_x\left\{n_Be^{i\chi\Delta}[C_2+C_0(1-\eta)]-(n_B+1)e^{-i\chi\Delta}\left[C_2-C_0\eta\right]\right\}.
\end{aligned}
\end{equation}
 We further introduced the coefficients
\begin{equation}
\label{eq:toyc0}
C_0 = \frac{4\sqrt{\lambda/\gamma}}{\sqrt{\pi}}e^{-4\frac{\lambda}{\gamma}}\sum_{n}^{\infty}\frac{(4\sqrt{\lambda/\gamma})^n}{(n+1)(n+1)!}H_n(2\sqrt{\lambda/\gamma})=
\frac{1}{2}\int_{0}^{1}\frac{dy}{y}\{{\rm erf}(2\sqrt{\lambda/\gamma})-{\rm erf}[2\sqrt{\lambda/\gamma}(1-2y)]\},
\end{equation}
\begin{equation}
\label{eq:toyc1}
\begin{aligned}
C_1 &= e^{-8\frac{\lambda}{\gamma}}\sum_{n}^{\infty}\frac{H^2_n(2\sqrt{\lambda/\gamma})}{2^{n+1}\pi(n+1)(n+1)!}=
\frac{\eta}{2}\int_{0}^{1}\frac{dy}{y}\{{\rm erf}(2\sqrt{\lambda/\gamma})-{\rm erf}[2\sqrt{\lambda/\gamma}(1-y)]\}\\&+\frac{1}{2\pi}\int_{0}^1dy\int_{0}^1dx\left\{\frac{e^{-4\frac{\lambda}{\gamma}\left[1+\frac{(1-y)^2}{1-y^2z^2}\right]}}{\sqrt{1-y^2z^2}}-\sqrt{4\pi\frac{\lambda}{\gamma}}e^{-4\frac{\lambda}{\gamma}(1-y+yz)^2}\left[1+{\rm erf}\left(2\sqrt{\lambda/\gamma}\frac{y(1-y)z+y^2z^2-1}{\sqrt{1-y^2z^2}}\right)\right]\right\},
\end{aligned}
\end{equation}

\begin{equation}
\label{eq:toyc2}
\begin{aligned}
C_2 &= e^{-8\frac{\lambda}{\gamma}}\sum_{n}^{\infty}\sum_{k=0}^n{n \choose k}\frac{(-1)^k(4\sqrt{\lambda/\gamma})^{n-k}}{2^{k+1}\pi(k+1)(n+1)!}H_n(2\sqrt{\lambda/\gamma})H_k(2\sqrt{\lambda/\gamma})=
\frac{\eta}{2}\int_{0}^{1}\frac{dy}{y}\{{\rm erf}[2\sqrt{\lambda/\gamma}(2y-1)]-{\rm erf}[2\sqrt{\lambda/\gamma}(y-1)]\}\\&+\frac{1}{2\pi}\int_{0}^1dy\int_{0}^1dx\left\{\frac{e^{-4\frac{\lambda}{\gamma}\left[1+\frac{(1-y)^2}{1-y^2z^2}\right]}}{\sqrt{1-y^2z^2}}-\sqrt{4\pi\frac{\lambda}{\gamma}}e^{-4\frac{\lambda}{\gamma}(1-y-yz)^2}\left[1-{\rm erf}\left(2\sqrt{\lambda/\gamma}\frac{y(1-y)z-y^2z^2+1}{\sqrt{1-y^2z^2}}\right)\right]\right\}.
\end{aligned}
\end{equation}

At zero counting field, we recover Eqs.~\eqref{eq:tW0} and Eqs.~\eqref{eq:tW1}, such that the steady state is $(1/2,1/2)^T$ as anticipated from the symmetry of the model. We then find the average power by
\begin{equation}
\label{eq:power}
P = -i\partial_{\chi}{\rm sum}\left\{[\tilde{W}_0(\chi)+\tilde{W}_1(\chi)/\gamma]\begin{pmatrix}
1/2\\ 1/2
\end{pmatrix}\right\}\bigg|_{\chi=0}=\Gamma\Delta\left[n_B(1-\eta)-(n_B+1)\eta\right]-\frac{\Gamma^2\Delta}{\gamma}(2n_B+1)[C_0 n_B-C_3],
\end{equation}
where the sum goes over all entries in the vector and we introduced $C_3 = C_1+C_2-\eta C_0$.

 After a long but straightforward calculation, one can verify that Eq.~(\ref{eq:power_classical_MD}) matches Eq.~(\ref{eq:power}) with the following mapping
\begin{equation}
\label{eq:map}
k = C_0,\hspace{1cm}m = -C_1, \hspace{1cm}z = C_0 \eta - C_2.
\end{equation}

\section{Quantum toy model}
\label{sec:quantum-toy}

Here we consider the Liouvillian
\begin{equation}
\label{eq:liouqtm}
\mathcal{L}(D)\hat{\rho} = -ig\cos(\Delta t)[\hat{\sigma}^\dagger+\hat{\sigma},\hat{\rho}]-i\theta(D)\Delta[|0\rangle\langle 0|,\hat{\rho}]-i[1-\theta(D)]\Delta[|1\rangle\langle 1|,\hat{\rho}],
\end{equation}
where $\hat{\sigma} = |0\rangle\langle 1|$. The power operator in this system is given by
\begin{equation}
\label{eq:powerop}
\hat{P}=\partial_t \hat{H}(t) = -g\Delta\sin(\Delta t)(\hat{\sigma}^\dagger+\hat{\sigma}),
\end{equation}
and the average power is determined by $\langle \hat{P}\rangle = {\rm Tr}\{\hat{P}\hat{\rho}\}$. Note that in general both the power operator as well as the density matrix are time-dependent objects.

\subsection{Analytical calculations}
\label{sec:quantum-toy-analytics}

To leading order in the separation of time-scales, the Liouvillian reduces to
\begin{equation}
\label{eq:l0qtm}
\mathcal{L}_0\hat{\rho}=-ig\cos(\Delta t)[\hat{\sigma}^\dagger+\hat{\sigma},\hat{\rho}].
\end{equation}
The steady state solution to this Liouvillian is simply the identity matrix which implies that $\langle\hat{P}\rangle=0$ to leading order in the separation of time-scales. This is related to the subtle interplay between $\gamma$ and $\lambda$ in the quantum regime. In the large bandwidth limit, the measurement strength $\lambda$ needs to go to zero in order for the separation of time-scales to be valid. Note that this trade-off is absent in the classical model as the time-scale of the system evolution is not determined by $\lambda$.

For the first order correction, we find
\begin{equation}
\label{eq:lcorrqtm}
\mathcal{L}_{\rm corr}\hat{\rho}=\frac{\Delta^2\ln(2)}{2}\mathcal{D}[\hat{\sigma}_z]\hat{\rho}-2\Delta gD_0\cos(\Delta t)(\hat{\sigma}^\dagger+\hat{\sigma}),
\end{equation}
with $D_0={}_2F_2(1/2,1/2;3/2,3/2;-4\lambda/\gamma)$.
The last term provides a source term for the off-diagonal elements of the density matrix. Such a source term may result in a density matrix that has negative eigenvalues. However, as long as the separation of time-scale assumption is justified, the term is small and this is not an issue.

The solution to the master equation $\partial_t\hat{\rho }=(\mathcal{L}_0+\lambda\mathcal{D}[\hat{\sigma}_z]+\mathcal{L}_{\rm corr}/\gamma)\hat{\rho}$ will tend to the periodic steady state
\begin{equation}
\label{eq:solqtm}
\hat{\rho}(t) = \mathbb{1}/2-(\hat{\sigma}^\dagger+\hat{\sigma})\frac{2\Delta g}{\gamma}D_0\frac{2\tilde{\lambda}\cos(\Delta t)+\Delta \sin(\Delta t)}{\Delta^2+4\tilde{\lambda}^2},
\end{equation}
where we introduced the effective dephasing parameter $\tilde{\lambda}=\lambda+\frac{\Delta^2\ln(2)}{2\gamma}$.
From the density matrix, it is straightforward to obtain the ensemble average of the power
\begin{equation}
\label{eq:powertime}
P(t)={\rm Tr} \{\hat{P}(t)\hat{\rho}(t)\} = \frac{4g^2\Delta^2}{\gamma}D_0\sin(\Delta t)\frac{2\tilde{\lambda}\cos(\Delta t)+\Delta \sin(\Delta t)}{\Delta^2+4\tilde{\lambda}^2}.
\end{equation}
If we average this over one period $\tau=2\pi/\Delta$, we recover the time-averaged power given in Eq.~(\ref{eq:quantum-power}) in the main text.

\subsection{Numerical calculations}
\label{sec:quantum-toy-numerics}

This section details the results of the numerical calculations for the quantum toy model. We begin by noting that the feedback Liouvillian in Eq.~\eqref{eq:liouqtm} is periodic in time. Assuming that the system reaches a periodic steady state, we expand the system-detector density operator as a Fourier series,
\begin{equation}
\hat{\rho}_t(D) = \sum_{q=-\infty}^\infty \hat{\rho}_q(D) e^{iq\Delta t},
\end{equation}
\noindent with time independent coefficients $\hat{\rho}_q(D) = \tau^{-1} \int_0^\tau dt \hat{\rho}_t(D) e^{-iq\Delta t}$, where $\tau=2\pi/\Delta$ is the driving period of the driving field. Equation (\ref{eq:MAIN-RESULT}) in the main text implies the following relation for the coefficients,
\begin{equation}
0 = \left[ \tilde{\mathcal{L}}(D) - iq\Delta \right]\hat{\rho}_q(D) + \hat{\mathcal{L}}\left[ \hat{\rho}_{q-1}(D) + \hat{\rho}_{q+1}(D) \right],
\end{equation}
\noindent with
\begin{equation}
\tilde{\mathcal{L}}(D)\hat{\rho} = -i[\hat{H}(D),\hat{\rho}] + \lambda \mathcal{D}[\hat{\sigma}_z]\hat{\rho} - \gamma\partial_D\mathcal{A}(D)\hat{\rho} + \frac{\gamma^2}{8\lambda}\partial_D^2\hat{\rho}, \hspace{2cm}
\hat{\mathcal{L}}\hat{\rho} = - \frac{ig}{2}[\hat{\sigma}_x,\hat{\rho}], 
\end{equation}
\noindent where $\hat{H}(D) = [1-\theta(D)]\Delta \ket{1}\!\bra{1} + \theta(D)\Delta\ket{0}\!\bra{0}$. By further assuming weak driving $g \ll \Delta$, and expanding the coefficients as $\hat{\rho}_q(D) \approx \hat{\rho}_q^{(0)}(D) + \frac{g}{\Delta} \hat{\rho}_q^{(1)}(D)$ gives the following relations,

\begin{subequations}
\begin{align}
0 &= [\Delta^{-1}\tilde{\mathcal{L}}(D)-iq]\hat{\rho}_q^{(0)}(D), \\
0 &= [\Delta^{-1}\tilde{\mathcal{L}}(D)-iq]\hat{\rho}_q^{(1)}(D) + g^{-1}\hat{\mathcal{L}}[\hat{\rho}_{q-1}^{(0)}(D)+\hat{\rho}_{q+1}^{(0)}(D)].
\end{align}
\label{eq:quantum-relations-fourier-space}
\end{subequations}
\noindent For Eq.~(\ref{eq:quantum-relations-fourier-space}a), only $q=0$ gives a non-zero solution and is given as a sum of steady state Ornstein-Uhlenbeck distributions,
\begin{equation}
\hat{\rho}_{q=0}^{(0)}(D) = \frac{1}{2}\sqrt{\frac{4\lambda}{\pi\gamma}} \left[ e^{-\frac{4\lambda}{\gamma}(D+1)^2}\ket{0}\!\bra{0} + e^{-\frac{4\lambda}{\gamma}(D-1)^2}\ket{1}\!\bra{1} \right].
\end{equation}
\noindent This implies that only $q=\pm 1$ provides non-zero solutions for Eq.~(\ref{eq:quantum-relations-fourier-space}b). These were found numerically using the method outlined in Sec.~\ref{sec:general-numerics}. The total density matrix $\hat{\rho}_t(D) = \hat{\rho}_{q=0}^{(0)}(D) + \frac{g}{\Delta}[ \hat{\rho}_{q=-1}^{(1)}(D)e^{-i\Delta t} + \hat{\rho}_{q=1}^{(1)}(D)e^{i\Delta t} ]$ was used to numerically calculate the power in Fig.~\ref{fig:toy-panel}(b), and the density matrix elements are plotted for $t=\tau$ in Fig.~\ref{fig:toy-panel}(c).

\end{document}